\documentclass[12pt]{article}
\usepackage{amsmath,amssymb,epsfig}
\usepackage{graphicx,subfigure}
\usepackage{verbatim,geometry}
\usepackage[normalem]{ulem}
\usepackage[usenames, dvipsnames]{color}
\usepackage{cite}
\numberwithin{equation}{section}

\makeatletter
\def\sla@#1#2#3#4#5{{%
  \setbox\z@\hbox{$\m@th#4#5$}%
  \setbox\tw@\hbox{$\m@th#4#1$}%
  \dimen4\wd\ifdim\wd\z@<\wd\tw@\tw@\else\z@\fi
  \dimen@\ht\tw@
  \advance\dimen@-\dp\tw@
  \advance\dimen@-\ht\z@
  \advance\dimen@\dp\z@
  \divide\dimen@\tw@
  \advance\dimen@-#3\ht\tw@
  \advance\dimen@-#3\dp\tw@
  \dimen@ii#2\wd\z@  \raise-\dimen@\hbox to\dimen4{%
    \hss\kern\dimen@ii\box\tw@\kern-\dimen@ii\hss}%
  \llap{\hbox to\dimen4{\hss\box\z@\hss}}}}
\def\slashed#1{%
  \expandafter\ifx\csname sla@\string#1\endcsname\relax
    {\mathpalette{\sla@/00}{#1}}%
  \else
    \csname sla@\string#1\endcsname
  \fi}
\makeatother

\makeatother

\newlength{\wth}
\newcommand{\startappendix}{
\setcounter{section}{0}
\renewcommand{\thesection}{\Alph{section}}}

\def\be{\begin{equation}}
\def\ee{\end{equation}}
\def\bear{\begin{eqnarray}}
\def\eear{\end{eqnarray}}

\def\half{{ \frac{1}{2} }}

\def\a{{\alpha}}

\def\p{{\partial}}

\def\e{{\epsilon}}

\def\vert{{|}}

\newcommand{\ba}{\begin{aligned}}
\newcommand{\ea}{\end{aligned}}

\begin{document}

\numberwithin{equation}{section}  
\allowdisplaybreaks  


%
%


\thispagestyle{empty}

\vspace*{-2cm} 
\begin{flushright}
EFI-11-20\\
KCL-MTH-11-14
\end{flushright}

\vspace*{0.5cm} 
\begin{center}
 {\LARGE \bf  On $G$-flux, M5 instantons, and $U(1)$s in F-theory\\}
\vspace*{1.0cm}
Joseph Marsano$^1$, Natalia Saulina$^2$ and Sakura Sch\"afer-Nameki$^3$

 \vspace*{1.0cm} 

{ \it$^1$ Enrico Fermi Institute, University of Chicago\\}
{ \it 5640 S Ellis Avenue, Chicago, IL 60637 USA\\}
{\tt marsano uchicago.edu \\}
\smallskip

{ \it$^2$ Perimeter Institute for Theoretical Physics \\}
{\it 31 Caroline St N., Waterloo, Ontario N2L 2Y5, Canada\\}
{\tt saulina theory.caltech.edu \\}
\smallskip

{ \it $^3$ Department of Mathematics, King's College, University of London \\ }
{ \it The Strand, WC2R 2LS, London, UK\\ }
{\tt ss299 theory.caltech.edu\\ }
 \vspace*{0.8cm} 
\end{center}


\begin{center} 
\textbf{Abstract} 
\end{center}

\noindent
Local aspects of singular F-theory compactifications for SUSY GUT model-building are fairly well understood in terms of Higgs bundles and their spectral data.  Several global issues remain, however, including a description of $G$-fluxes, which are key to constructing chiral matter and stabilizing moduli, and the global realization of $U(1)$ symmetries that can forbid phenomenologically unfavorable couplings.  In this paper, we sharpen our earlier proposal for describing $G$-fluxes through ``spectral divisors'' and introduce a distinguished ``Tate divisor'', which can be used to describe both $G$-flux and $U(1)$s when present.  As an application, we give a general discussion of M5-instanton contributions in the presence of $G$-flux and exemplify this in a concrete example, where we comment on the ability of instanton induced superpotential couplings to stabilize K\"ahler moduli.




\newpage
\tableofcontents


\section{Introduction and Summary}

While the rules for local{\footnote{We refer to  ``local" in this context as the complete description of the 8d SYM theory on 7-branes in terms of Higgs bundle data, which does include information on the local geometry and monodromies (also sometimes refered to as ``semi-local").}}  model-buiding in F-theory GUTs are becoming fairly well understood \cite{Donagi:2009ra,Marsano:2009gv,Dudas:2010zb,Marsano:2010sq,Ludeling:2011en,Dolan:2011iu}, any such model still carries intrinsic assumptions about the global completion.   
For starters, local models describe only a local notion of the fluxes required to generate chiral matter.  Further, local models often rely on $U(1)$s which have become massive through a St\"uckelberg mechanism \cite{Marsano:2009gv,Marsano:2009wr,Dolan:2011iu} to solve important phenomenological problems.  These $U(1)$s are intrinsically global in nature, though, so their presence or absence depends on details of the global completion
 \cite{Hayashi:2010zp}.

These two points have received some attention in the past \cite{Hayashi:2010zp,Grimm:2010ez,Marsano:2010ix} but they do not represent the only bulk physics to which local models must appeal.  In many models \cite{Marsano:2008jq,Heckman:2008qt,Tatar:2009jk,Marsano:2009wr,Dudas:2009hu,Dolan:2011iu} the presence of GUT-singlets must be assumed for a variety of reasons, whether it be to generate neutrino mass, break supersymmetry, or lift unwanted exotic particles.  Crucial in all such models are the superpotential couplings involving the singlets, which are typically assumed to be generated by some bulk physics.  Here, bulk physics usually means one or more M5-instantons.  Of course M5-instantons are important for more than just singlet physics; the issue of moduli stabilization, which will involve M5-instantons in a crucial way, is lurking behind everything that local model-builders do.

Our objective in this paper is to make some progress toward understanding these global issues in order to build continually improving global completions of local models.  We begin with a discussion of the two important  constructs that underlie all global models: $G$-flux and $U(1)$ symmetries.  This discussion aims to extend and clarify that of \cite{Marsano:2010ix} and clearly state how $U(1)$ symmetries can be understood within the ``spectral divisor" formalism through the introduction of a distinguished object that we refer to as the ``Tate divisor". $U(1)$s have received careful treatment also in \cite{Grimm:2010ez} and we believe our approach is consistent with the results contained therein.
 Our construction extends the spectral cover description of Higgs bundles and $G$-fluxes in the local setup to the full Calabi-Yau fourfold.  It is motivated from heterotic/F-theory duality \cite{Friedman:1997yq,Curio:1998bva,Andreas:1999ng,Donagi:2008ca,Donagi:2008kj,Hayashi:2009ge} and is compatible with the spectral cover in that context whenver such a dual Heterotic description exists \cite{Marsano:2010ix}.  Along the way, we make comments on $U(1)$ fluxes, the D3-brane tadpole, and flux quantization. 

We then turn to a discussion of M5-instantons where our aim is to study the conditions for which the couplings that they generate are nonzero.  There is a vast literature on M5-instantons detailing many of these issues, {including recent work focused on the connection to F-theory and the relation to D3-brane instantons in type IIB \cite{Heckman:2008es,Marsano:2008py,Cvetic:2009ah,Blumenhagen:2010ja,Cvetic:2010rq,Donagi:2010pd,Cvetic:2010ky,Grimm:2011dj}}.
We pay particular attention to recasting the fermi zero mode computation in terms of cohomologies on divisors in the base $B_3$ of $Y_4$ and the interplay of M5-instantons with the $G$-fluxes that we construct with spectral divisors as in \cite{Marsano:2010ix}.  With this knowledge, we turn to geometries based on the threefold of \cite{Marsano:2009ym}, specify which divisors support M5s that generate nonzero couplings, and comment on the implications for moduli stabilization in those models.  We also take this opportunity to present a cleaner description of the threefold in \cite{Marsano:2009ym}.  Before getting to the details, let us summarize our general approach to these issues as well as some of the results. 

\subsection{Approach to $G$-flux and $U(1)$ symmetries}

The constructs of $G$-flux and $U(1)$ symmetries are not completely unrelated because each admits a similar geometric description.  Consider, for instance, an elliptically fibered Calabi-Yau fourfold $Y_4$ with a surface of $SU(5)_{\rm GUT}$ singularities.  The $G$-fluxes that generate chiral matter are $(2,2)$-forms in $Y_4$ that have exactly 1 leg on the torus, which is to say that they can be thought of as (linear combinations of) holomorphic surfaces that are effectively orthogonal to all horizontal and vertical divisors.  If $Y_4$ were smooth this condition would tell us that $G$ is completely trivial because it would integrate to zero over every holomorphic surface but $Y_4$  is crucially not smooth.  The resolution $\tilde{Y}_4$ will contain holomorphic surfaces that do not sit inside the preimage of horizontal or vertical divisors under the blow-down map $P:\tilde{Y}_4\rightarrow Y_4$.  These include the matter surfaces that one obtains when a curve of singularities is resolved{\footnote{Of course these also include surfaces that sit inside the divisors that one obtains from the resolution of surfaces of singularities.}}.  Curves of singularities support matter fields that descend from wrapped M2-branes and our $G$-flux should have nonzero integral over some of the matter surfaces that result from resolving those singularities.  Any nontriviality in the $G$-flux, then, is crucially tied to the resolution. 

The issue with $U(1)$ symmetries is very similar.  $U(1)$s that couple to our charged matter fields, which themselves originate from wrapped M2-branes, come from the reduction of the M-theory 3-form $C_3$ on harmonic $(1,1)$-forms $\omega$
\begin{equation}C_3=A_1\wedge \omega \,.
\end{equation}
When we compactify M-theory on $Y_4$, this gives a 3-dimensional gauge field on $\mathbb{R}^{2,1}$.  To ensure that this carries over to a 4-dimensional gauge field in the F-theory limit, where the volume of the elliptic fiber is scaled to zero, the $(1,1)$-form $\omega$ must be chosen to have 1 leg on the elliptic fiber.  If we wish to think of $\omega$ as a divisor in $Y_4$, then, this condition means that $\omega$ cannot be horizontal or vertical.  If $Y_4$ were smooth, this would force $\omega$ to be trivial so once again any nontriviality in $\omega$ is intrinsically tied to the singularities.  This is to be expected because we anticipate that $\omega$ has nonzero integral over degenerate cycles.  This is how 4-dimensional states from wrapped M2-branes manage to couple to $U(1)$ vector bosons in the first place.

In each of these cases we seek a simple geometric description, whether it be a holomorphic surface to specify $G$ or a divisor to specify the $(1,1)$-form $\omega$ that we need to get a $U(1)$.  Unfortunately, these concepts must be refined when the geometry is singular and all of the nontriviality is contained in that refinement.  In this paper, as in previous work \cite{Marsano:2010ix}, we adopt the approach that the physics on a singular Calabi-Yau fourfold $Y_4$ should be specified by starting with a particular smooth resolution $\tilde{Y}_4$ and taking a singular limit.
 This is essentially what we do when assigning gauge groups and charged matter to ADE singularities anyway.  We specify the physics by saying that we start with a resolution $\tilde{Y}_4$ in which the ADE singularities are resolved by small resolutions, i.e. growing 2-cycles (as opposed to, say, deforming).  We can wrap M2-branes on the finite volume 2-cycles in $\tilde{Y}_4$ where their physics is clear and then we see that these states become massless in the singular limit.  The physical D3-brane charge should also include the geometric contribution from the resolved geometry $\tilde{Y}_4$ as computed in several examples over the past couple of years \cite{Blumenhagen:2009yv,Grimm:2009yu}.  In the same way, we should understand backgrounds for $C_3$, including $G$-fluxes and reductions of $C_3$ that yield $U(1)$s, in the resolved geometry $\tilde{Y}_4$.  $G$-flux of the type that we need to induce chirality should be describable as a holomorphic surface in $\tilde{Y}_4$ and $U(1)$s should be describable as ordinary divisors in $\tilde{Y}_4$.  In each case, the ``1 leg on the torus" condition just tells us that the corresponding surface or divisor is orthogonal to everything that survives the singular limit.

To describe $G$-flux and $U(1)$ symmetries, then, it is crucial to specify a resolution $\tilde{Y}_4$.  
We actually do not describe the resolution explicitly but rather, following ideas from Heterotic/F-theory duality, we use $dP_9$-fibrations to do the job.  The basic strategy is to take advantage of the fact that we are interested in small resolutions which should be describable entirely within the local geometry near a surface of singularities, $S_{\rm GUT}$, which we typically assume to be of $A_4$ type.  Because the resolution does not care about what happens beyond the local geometry, we can excise it from $Y_4$ and insert it into a different fourfold $\hat{Y}_4$ 
that has the structure of a $dP_9$-fibration.  This allows us to describe the vanishing cycles in terms of the homology of $dP_9$ and, in particular, to think about the singularity structure in terms of the Mori cone of the $dP_9$ fiber as one moves along the base $S_{\rm GUT}$.  The geometry of $dP_9$ tells us about the structure of small resolutions, which simply grow back the zero volume homology cycles that sit in the Mori cone, and allows us to specify holomorphic surfaces and divisors in $\hat{Y}_4$ that interact in the right way with resolved cycles.  We can then restrict these to the local geometry and, in so doing, obtain noncompact surfaces and divisors that we can try to globally extend after putting everything back into $Y_4$.  

A procedure based on $dP_9$ is a natural way to extend the ideas of F-theory/Heterotic duality beyond the realm of K3-fibered Calabi-Yau fourfolds and is at the heart of the approach in \cite{Marsano:2010ix}.  The spectral divisor in that work was just a tool for understanding holomorphic divisors and surfaces in the resolved geometry in terms of objects that we can describe in the singular one.  In this work, we extend these ideas by introducing a distinguished spectral divisor that allows for the engineering of $U(1)$ symmetries.  We also comment on how the Higgs bundle picture naturally emerges and discuss briefly some issues related to the counting of GUT-singlets, the D3-brane tadpole, and flux quantization.  Note that the $dP_9$ approach only works, of course, when the surface of singularities contains no structure larger than $E_8$.  Gauge groups that are not of this type require cycles that simply do not fit into the homology of $dP_9$, as reflected by the fact that a local geometry capable of engineering such gauge groups cannot be embedded into a $dP_9$-fibration.

As we were finalizing this work  \cite{Esole:2011sm} appeared, where small resolutions of singular Calabi-Yau fourfolds with an $SU(5)$ singularity are studied in general. There are several small resolutions, which are birationally equivalent, and it would be interesting to put our analysis of $G$-fluxes and $U(1)$s in this context. 


\subsection{M5-instantons}

In the second part of this paper we turn to the study of M5-instantons, by which we really mean those Euclidean M5s in M-theory compactified on $Y_4$ that descend to D3-brane instantons in the F-theory limit.  M5s of this type wrap vertical divisors of the form $\pi^*S_2$ with $S_2$ a divisor in the base $B_3$ of $Y_4$ and $\pi: Y_4 \rightarrow B_3$ is the elliptic fibration.  In this setting, we focus on the two things that can destroy an M5s ability to contribute a superpotential coupling: fermi zero modes and $G$-flux.  Because our M5s are really describing type IIB D3s wrapping $S_2$\footnote{{For a review of D3-instantons in type IIB orientifolds see \cite{Blumenhagen:2009qh}}.}, one expects that the computation of fermi zero modes should be related to how ``moveable" $S_2$ is inside $B_3$, that is to cohomologies of the normal bundle $N_{S_2|B_3}$.  We work out this relation using the Leray spectral sequence and then use those results to determine all surfaces $S_2$ in the threefold of \cite{Marsano:2009ym} that can support an instanton with the right fermi zero mode structure.

We then discuss the interplay of $G$-flux with M5-instantons.  If the restriction of $G$ to an M5 is nontrivial, its effect is to introduce a source term for the chiral 2-form $b_2$.  This source can sometimes be effectively canceled by introducing wrapped M2-branes ending on the M5 and, in those cases, the M5 generates a charged coupling involving the wrapped M2-brane states.  One can see on general grounds that such a coupling will not only be invariant under any $U(1)$s that were engineered but more generally that it will be invariant under bulk gauge transformations of the M-theory 3-form $C_3$.  We believe this should incorporate the selection rules alluded to in \cite{Grimm:2011dj} involving ``massive $U(1)$s" at the KK scale.  When the source cannot be canceled by wrapped M2-branes, we believe the superpotential coupling is forced to vanish{\footnote{The absence of wrapped branes that can cancel the source amounts to the statement that there are no classical solutions to the equations of motion for $b_2$ in the presence of the source that are properly quantized.  There is a factor of $\frac{1}{2}$ that we discuss which leads to a scenario where wrapped branes cannot cancel the source.  It is possible that some object exists that can do the job but we do not see a candidate at the moment.}}.

Using the spectral divisor description of $G$-fluxes, we are able to argue that fluxes used to generate chirality in F-theory GUT models generically have trivial restriction to M5-instantons so they generate ``uncharged" couplings that can play a role in moduli stabilization when the fermi zero mode structure is correct.  This is actually easy to see without a detailed computation.  The $G$-fluxes for chirality are engineered so that they only have nonzero integral over matter surfaces obtained by combining a curve of singularities $\Sigma_{sing,i}$ with the curve $C_i$ that degenerates there.  In order for $G$ to have nontrivial restriction to the M5, then, it must contain a full curve of singularities in $Y_4$ and hence contain a full matter surface in its entirety in $\tilde{Y}_4$; otherwise, $G$ would integrate to zero over all divisors in the M5.  Any curve of singularities, though, will generically meet an M5 in isolated points in $Y_4${\footnote{Of course, the M5 will contain curves of singularities $\Sigma_{sing}'$ along which it intersects surfaces of singularities like $S_{\rm GUT}$.  The only $G$-flux that would integrate nonzero under the surface obtained by resolving the singularity over $\Sigma_{sing}'$ will be one that corresponds to a worldvolume flux.  In F-theory GUT models we typically do not  turn on any worldvolume fluxes except for hypercharge flux, which is globally trivial.  As the intersection of the M5 with a surface $S_{\rm GUT}$ of singularities in $Y_4$ is isomorphic to a curve in $S_{\rm GUT}$ that is nontrivial in the base $B_3$ of $Y_4$, the hypercharge flux will integrate to zero over it.}}.  The restriction of $G$ to such an M5 will be trivial and the coupling generated, if any, will be ``uncharged" in the sense that it will involve only moduli and not any fields that arise from wrapped M2s.

The upshot of this analysis is that vertical divisors of elliptically-fibered fourfolds based on the threefold geometry of \cite{Marsano:2009ym} that we identify as having the right zero mode structure will have no problem generating uncharged couplings if the only $G$-fluxes we introduce are spectral divisor fluxes for inducing a chiral spectrum.  This gives an indication of whether we have any right to expect K\"ahler moduli stabilization to be possible in geometries constructed from the threefold of \cite{Marsano:2009ym}.  Note that to combine K\"ahler moduli stabilization with complex structure moduli stabilization we must assume that the latter does not require any $G$-fluxes that restrict nontrivially to the M5 instantons that are used for the former.  This is not a new problem and it is somewhat canonical to assume that complex structure moduli stabilization does not interfere with K\"ahler moduli stabilization in this way.  Such an assumption should not be made cavalierly, though; it is an important one whose validity cannot be taken for granted.
Any claims to actually achieve moduli stabilization in an F-theory GUT model must justify it.  To be sure, we make no such claims in the present paper.

\subsection{Outline}

The remainder of this paper is organized as follows.  In section \ref{sec:Tate} we extend the spectral divisor formalism of \cite{Marsano:2010ix} to describe $U(1)$s as well as $G$-fluxes.  We also comment on the connection to Higgs bundles, the D3-brane tadpole, and flux quantization.  In section \ref{sec:M5}, we turn to the study of M5-instantons and discuss the two ways that their contribution to the superpotential can be ruined: fermi zero modes and $G$-flux.  Finally, in section \ref{sec:application} we apply  these ideas in some generality to geometries based on the threefold of \cite{Marsano:2009ym}.  We identify divisors in the threefold that form the base of vertical divisors in an elliptically fibered Calabi-Yau fourfold that can lead to nontrivial M5 instanton corrections.  We also verify that the $G$-flux needed to generate chirality will restrict trivially to the worldvolumes of those M5s.  
Appendix \ref{app:ThreeFold} contains a cleaner description of the threefold geometry that we use than the one contained in \cite{Marsano:2009ym}.  Because this description does not  utilize any flop transitions we believe it to be more transparent.

\section{$U(1)$s, $G$-flux, and the ``Tate Divisor"}
\label{sec:Tate}


In this section, we review and extend the formalism described in \cite{Marsano:2010ix} for describing $(2,2)$ $G$-fluxes in F-theory GUTs that generate chiral spectra.  We would like to emphasize a few things that were not addressed in \cite{Marsano:2010ix}.  The first is a description of how the spectral divisor formalism can be used to take the apparent $U(1)$s of a local F-theory GUT and explicitly engineer these gauge bosons in a global extension{\footnote{That this issue is subtle was emphasized in \cite{Hayashi:2010zp} while a focused study of $U(1)$s in global F-theory models, with which we believe our discussion is consistent, can be found in \cite{Grimm:2010ez}.}}.  To explicitly realize $U(1)$s we will be lead to consider a distinguished object that we term the ``Tate divisor", which is a special example of a type of object referred to as a ``spectral divisor" in \cite{Marsano:2010ix}{\footnote{As described in \cite{Marsano:2010ix}, a spectral divisor in our singular Calabi-Yau fourfold is one that contains important singularities and whose limiting behavior near those singualrities causes it to behave in a favorable way upon resolution.  The ``Tate divisor" is a distinguished example of such a divisor that is suitable for describing not only $G$-fluxes but also $U(1)$s, as we shall see.}}.  In addition to allowing an explicit description of $U(1)$s, the ``Tate divisor" also helps to obtain an explicit description of $U(1)$ flux following ideas of \cite{Marsano:2010ix}.  After this, we turn to a discussion of chiral spectra, whose determination we formulate in a language that is more amenable to the study of GUT singlets that localize on curves that do not sit inside the GUT surface (or any other surface above which the elliptic fibration exhibits non-Abelian singularities).  Finally, we make some comments about the D3-brane tadpole including a general argument that the contribution from our $G$-fluxes arises only from surfaces of singularities in accord with a conjecture of \cite{Grimm:2009yu}{\footnote{More specifically, \cite{Grimm:2009yu} checks that the difference in the Euler character of several resolved fourfolds and the naive computation of the Euler character based on data of the singular fourfold only seems to get contributions from surfaces of singularities.  As authors of \cite{Grimm:2009yu} have suggested, this is indicative of flux-induced tadpoles getting contributions only from surfaces of singularities as well.}}.  The extension of ideas that we describe here to surfaces of exceptional type singularities should be straightforward and we have some comments on this.  Further generalizations are beyond the scope of this paper but would be an interesting topic of future study.

\subsection{Geometric Setup and $U(1)$s}

Our basic setup is an elliptically fibered Calabi-Yau fourfold, $Y_4$, that exhibits a surface of $SU(5)_{\rm GUT}$ singularities{\footnote{There may be hidden sectors consisting of additional surfaces of singularities but we ignore this possibility for now, and will return to this in the context of M5 instantons. }}.
We let $\pi$ denote the elliptic fibration
\begin{equation}\pi:Y_4\rightarrow B_3\end{equation}
and write the defining equation for $Y_4$ in the ``Tate form"
\begin{equation}y^2 = x^3 + a_0z^5 + a_2z^3x + a_3z^2y + a_4zx^2 + a_5xy \,.
\label{tateform}\end{equation}
Here, $z$ is the holomorphic section on $B_3$ whose vanishing defines the surface $S_{\rm GUT}$ and the $a_m$'s are sections of the bundles ${\cal{O}}((m-6)K_{B_3}+(m-5)S_{\rm GUT})$.  This geometry exhibits an $SU(5)_{\rm GUT}$ singularity along $z=0$ \cite{Bershadsky:1996nh} with enhancements in singularity type to $SO(10)$ where $z=a_5=0$ and $SU(6)$ where $z=a_0a_5^2 -a_2a_3a_4+a_3^2a_4=0$.  The divisor $z=0$ supports the degrees of freedom of an ${\cal{N}}=1$ supersymmetric gauge theory with gauge group $SU(5)$ while the curves of $SO(10)$ and $SU(6)$ enhancement support charged chiral matter.  The string theoretic origin of that matter is well understood: matter fields arise from M2-branes wrapping the extra vanishing cycles or, equivalently, the $(p,q)$ strings that become massless there in the type IIB language.

In this setting, both $U(1)$ symmetries and the $G$-fluxes responsible for chiral spectra are somewhat subtle to describe.  This is because both are sensitive to the monodromic structure of the geometry near $z=0$ (among other things), where the $SU(5)_{\rm GUT}$ degrees of freedom live.  We focused on $G$-fluxes in \cite{Marsano:2010ix} so we will place a greater emphasis on $U(1)$s in the present paper.

One of the nice features of $U(1)$s from a phenomenological perspective is that they can distinguish different types of 4-dimensional fields that sit in the same $SU(5)_{\rm GUT}$ representation.  We are usually interested in distinguishing $\mathbf{\overline{5}}$'s (so that left-handed leptons do not participate in the same couplings as down type Higgs doublets) but, for simplicity of presentation, we focus instead on $\mathbf{10}$'s that localize on the curve $z=a_5=0$.  This curve, $\Sigma_{\mathbf{10}}$, is the locus of $SO(10)$ singularities and, if $a_5$ is chosen appropriately, may split into multiple components.  The type of $U(1)$ we are interested in should couple differently to $\mathbf{10}$'s associated to the different components so, to think about $U(1)$s geometrically, we ask a related question: is there a way to distinguish (or tell if it is possible to distinguish) between the 2-cycles that degenerate along different components of $\Sigma_{\mathbf{10}}$ when it is reducible?  In trying to distinguish these 2-cycles geometrically we will end up building the $U(1)$ that we want.

As reviewed in the introduction, $U(1)$s in F-theory originate from harmonic 2-forms, which integrate to zero over the fiber class, $F$, and any curve that sits inside the section of the elliptic fibration.  
Given such a 2-form in our Calabi-Yau fourfold, we can reduce the M-theory 3-form $C_3$ on it to obtain a 3-dimensional gauge field that becomes a 4-dimensional gauge field in the F-theory limit (where the elliptic fiber is shrunk to zero volume while holding $\tau$ fixed).  
The $U(1)$ charge of chiral fields that descend from M2-branes wrapped on a cycle $C$, then, is simply the integral of $\omega$ over $C$ 
\begin{equation}\label{m2charge}
q_{C}\sim\int_{C}\omega \,.
\end{equation}
The cycles $C$ that house chiral matter are holomorphic so we will want $\omega$ to be a $(1,1)$-form.  Alternatively, we specify $\omega$ by a dual divisor $D_\omega$ in the resolved geometry, so that
\be
q_C \sim D_\omega \cdot C \,.
\ee
If two curves $C_1$ and $C_2$ that support wrapped M2s are homologously distinct, it should be possible to construct a divisor $D_\omega$ such that
\be
D_\omega\cdot C_1\ne D_\omega\cdot C_2 \,.
\ee
 This can be used to construct a $U(1)$ with respect to which the fields from $C_1$ and $C_2$ carry distinct charges.

\subsection{Local geometry and $dP_9$}

Because $U(1)$ charges involve integrations over vanishing 2-cycles, the existence of a $U(1)$ that distinguishes charged fields is closely tied to how singularities are resolved.  This can be quite complicated in general so our strategy will be to take a somewhat indirect route motivated by the cylinder map of Heterotic/F-theory duality \cite{Friedman:1997yq,Curio:1998bva,Donagi:2008ca,Donagi:2008kj,Hayashi:2009ge}.  Rather than explicitly resolving all singularities, we will embed our local geometry into a different fourfold $\hat{Y}_4$ in which the singularity structure can be related to the geometry of a family of $dP_9$ surfaces.  
 The advantage of embedding our local geometry into a family of $dP_9$'s  is that we can describe the singularity structure in terms of $dP_9$ homology cycles that degenerate above various loci in $S_{\rm GUT}${\footnote{Of course this limits the type of cycles we can consider but there is no problem for phenomenologically relevant examples in which all singularities that we deal with embed into $E_8$.}}.

To proceed, then, consider the local geometry
\begin{equation}y^2 = x^3 + b_0z^5 + b_2z^3x + b_3z^2y + b_4zx^2 + b_5xy\end{equation}
where the $b_m$'s are just the restrictions of the global holomorphic sections $a_m$ in \eqref{tateform} to $S_{\rm GUT}$
\begin{equation}b_m = a_m|_{S_{\rm GUT}}\,.
\end{equation}
This is the local model for the geometry near $S_{\rm GUT}$ and its defining data admits a well-known mapping to Higgs bundle data of the 7-brane worldvolume gauge theory \cite{Donagi:2008ca,Donagi:2008kj,Hayashi:2009ge,Donagi:2009ra,Donagi:2011jy}.  It is a trivial matter to embed this local geometry into a K3-fibered Calabi-Yau fourfold $\hat{Y}_4$ by embedding $S_{\rm GUT}$ into the threefold $\hat{B}_3=\mathbb{P}({\cal{O}}\oplus N)$ with $N$ the normal bundle of $S_{\rm GUT}$ inside $B_3$.  By construction, our surface $S_{\rm GUT}$ has an identical normal bundle in $\hat{B}_3$ as in $B_3$ and all of the holomorphic sections $b_m$ can be trivially extended to global ones in $\hat{B}_3$ that can be used to define $\hat{Y}_4${\footnote{It may be that $\hat{Y}_4$ has some rather sick singularities far from $S_{\rm GUT}$.  This will not have any affect on what we say here, though, as our interest is only in small resolutions of singularities along $S_{\rm GUT}$.}}.

With this K3-fibered Calabi-Yau we can utilize Heterotic/F duality. Indeed, the spectral divisor formalism for $G$-fluxes \cite{Marsano:2010ix} is a strategy to apply ideas of the cylinder map in a more general setting 
\cite{Friedman:1997yq,Curio:1998bva,Donagi:2008ca,Hayashi:2008ba,Donagi:2008kj,Donagi:2009ra}.  We are not interested in Heterotic/F duality here per se, but rather in the stable degeneration limit \cite{Morrison:1996na,Morrison:1996pp} of $\hat{Y}_4$ so that our local geometry becomes embedded into a $dP_9$ fibration, which we think of as a family of $dP_9$'s parametrized by $S_{\rm GUT}$.  
We can explicitly present the family as 
\begin{equation}y^2 = x^3 + f_4z^4 + g_6z^6 + w\left(b_0z^5 + b_2z^3x + b_3z^2y + b_4zx^2 + b_5xy\right) \,,
\label{dp9fib}\end{equation}
where $w$ and $z$ are homogeneous coordinates on the $\mathbb{P}^1$ fiber of $\hat{B}_3${\footnote{Because the power of $w$ out front is only 1 this does not define a Calabi-Yau fourfold.  Rather, it is only `half' of the Calabi-Yau fourfold $\hat{Y}_4$ that we can split off in the stable degeneration limit.  It has the advantage that if the local geometry of $Y_4$ is well behaved then the $dP_9$-fibration is also well-behaved in that there are no `sick' singularities at $w=0$.  Even though the $dP_9$-fibration is not itself Calabi-Yau, it is a fine place to study the small (crepant) resolutions we need to desingularize $Y_4$.  As described in the introduction, we could skip the K3-fibration altogether and simply start by embedding our local geometry into the $dP_9$-fibration \eqref{dp9fib}.}}.  

In presenting the $dP_9$ as an elliptic fibration over $\mathbb{P}^1$ with section, we have distinguished two curve classes.  These are the base, $e_9$, and the elliptic fiber, $F$, which is a representative of the anti-canonical class.  Additional cycles that degenerate correspond to $-2$ curves that fail to intersect either $e_9$ or $F$.  
Elements of $H_2(dP_9,\mathbb{Z})$ with this property are referred to as roots and, following \cite{Hayashi:2008ba}, we label this collection by $R_8$
\begin{equation}
R_8 = \{C\in H_2(dP_9,\mathbb{Z})|C\cdot F=C\cdot e_9=0,\,\,C^2=-2\} \,.
\end{equation}
The intersection matrix on $R_8$ is equivalent to ($-1$ times) the Cartan matrix of $E_8$ so the roots $C$ are naturally identified with elements of the $E_8$ root lattice.  One inconvenient feature of the $C$'s is that they do not have effective holomorphic representatives in generic $dP_9$ surfaces (i.e. they do not sit in the Mori cone).  Of course, these $dP_9$'s are somewhat special in that all of them exhibit $SU(5)_{\rm GUT}$ singularities.  The 4 roots of $SU(5)_{\perp}$ are therefore in the Mori cone throughout the family{\footnote{ This can be achieved in the standard picture of $dP_9$ as $\mathbb{P}^2$ blown up at 9 points by taking the limit as several of the blown up points move to nongeneric locations, such as sitting on top of one another or sitting on a common line.}} so the monodromy group as we move in the family must sit inside the Weyl group of the $SU(5)_{\perp}$ commutant of $SU(5)_{\rm GUT}$ inside $E_8$, which is the symmetric group on five objects.


\subsection{From $dP_9$ to $U(1)$s}
\label{subsec:dp9U1}

The $dP_9$ fibers over matter curves like $\Sigma_{\mathbf{10}}$ exhibit an additional holomorphic curve corresponding to the new vanishing cycle.  We can think of this as coming from a new root that enters the Mori cone.  Our task of distinguishing different kinds of $\mathbf{10}$'s boils down to determining whether the new effective root on one component of $\Sigma_{\mathbf{10}}$ is related by monodromy to the new effective root on another component as we move in the family.  The easiest way to do this would be to simply follow one of the roots through the family.  This cannot be done easily because the root exits the Mori cone as soon as we move off of $\Sigma_{\mathbf{10}}$.  Fortunately, however, the roots are in 1-1 correspondence with elements of a second distinguished set
\begin{equation}I_8 = \{\ell\in H_2(dP_9,\mathbb{Z})|\ell\cdot F=1,\,\,\ell\cdot e_9=0,\,\,\ell^2=-1\}
\label{elldefs}\end{equation}
according to the identification
\begin{equation}\ell = (F+e_9)-C\,.
\label{IRmap}\end{equation}
Elements of $I_8$ are precisely the exceptional lines of $dP_9$ that miss the section $e_9$.  They are in the Mori cone of generic $dP_9$'s  so they have nice effective holomorphic representatives throughout the family.  If we want to follow the behavior of a particular root as we move along $S_{\rm GUT}$, then, it is often easier to look at what happens to the effective line from $I_8$ that is dual to that root in the sense of \eqref{IRmap}.  

We are interested in a particular subset of the roots that correspond to $\mathbf{10}$'s of $SU(5)_{\rm GUT}$.  The group theoretic decomposition of the $E_8$ adjoint under \begin{equation}
\ba
E_8 & \quad \rightarrow \quad SU(5)_{\rm GUT}\times SU(5)_{\perp}\cr
\mathbf{248} &\quad \rightarrow \quad (\mathbf{24},\mathbf{1})\oplus (\mathbf{1},\mathbf{24})\oplus \left[(\mathbf{10},\mathbf{5})\oplus\text{cc}\right]\oplus\left[(\mathbf{\overline{5}},\mathbf{10})\oplus\text{cc}\right]
\ea
\end{equation}
tells us that, modulo the action of $SU(5)_{\rm GUT}$ roots, there are five such roots, transforming as a fundamental of $SU(5)_{\perp}$, that can move into the Mori cone above a given component of $\Sigma_{\mathbf{10}}$.  To see how their homological classes mix in the family, we study instead the behavior of the 5 exceptional lines $\ell_i$ dual to those roots in the sense of \eqref{IRmap}.
Given the form of the $dP_9$ fibration \eqref{dp9fib}, we can describe the union of these five exceptional lines (fibered over $S_{\rm GUT}$)  by the divisor ${\cal{C}}$ \cite{Hayashi:2008ba}{\footnote{When restricted to any single $dP_9$ of the family, ${\cal{C}}$ describes a union of five exceptional lines.  Because it is irreducible for a generic family with $SU(5)_{\rm GUT}$ singularity at $z=0$, these five exceptional lines are mixed by the generic $S_5$ monodromy group and hence are precisely the $\ell_i$'s that are dual to the $\mathbf{10}$ roots in the sense of \eqref{IRmap} provided we take a proper transform in the $dP_9$.}}
\begin{equation}
{\cal{C}}:\quad b_0z^5 + b_2z^3x + b_3z^2y+b_4zx^2+b_5xy  =0 \,.
\label{ellis}\end{equation}
To ensure that some $\mathbf{10}$'s are distinct from others, what we need is for the divisor \eqref{ellis} to split into multiple components in our $dP_9$-fibration.  If this happens, some exceptional lines will not mix with others in the fibration.  We will have two distinct sets of lines $\{\ell_i\}$ and $\{\ell_a\}$ and, correspondingly, two distinct sets of roots, $\{C_i\}$ and $\{C_a\}$.

Along with guaranteeing that the $\{C_i\}$ and $\{C_a\}$ do not mix, the splitting of our exceptional lines into the sets $\{\ell_i\}$ and $\{\ell_a\}$ also provides a means for distinguishing them.
For concreteness, let us suppose that there are $n$ roots in the group $\{C_i\}$ and $m$ in the group $\{C_a\}$.  In that case, consider the following element of $H_2(dP_9,\mathbb{Z})$
\begin{equation}
w = m\sum_{i=1}^n\ell_i - n\sum_{a=1}^m\ell_a \,.
\end{equation}
This is a nontrivial element of $H_2(dP_9,\mathbb{Z})$ that distinguishes roots in the sets $\{C_i\}$ and $\{C_a\}$ via the intersection relations
\begin{equation}
w\cdot C_i = m \,,\qquad w\cdot C_a = -n \,,
\label{wCint}\end{equation}
which follow from the local intersection data between the exceptional lines $\ell_i$ and roots $C_i$ ($i=1,\ldots,5$) in $dP_9$
\begin{equation}\ell_i\cdot C_j = 1+\delta_{ij} \,.
\label{localint}\end{equation}
The action of $w$ therefore reduces to that of a Cartan generator of $SU(5)_{\perp}$
\begin{equation}w\leftrightarrow \text{diag}(m,m,\ldots,m,-n,-n,\ldots,-n) \,.
\end{equation}
When the divisor \eqref{ellis} splits into two components ${\cal{C}}^{(m)}$ and ${\cal{C}}^{(n)}$ with $m$ and $n$ sheets, respectively, the combination $w$ is invariant under the monodromy action of our family so that it fibers over $S_{\rm GUT}$ to yield a nontrivial divisor given by the linear combination $n{\cal{C}}^{(m)}-m{\cal{C}}^{(n)}$.  This divisor, in turn, provides a $(1,1)$-form that can be used to obtain a $U(1)$ gauge field that couples differently to M2-branes wrapping the roots $\{C_i\}$ and $\{C_a\}$ according to the group theoretic intersection \eqref{wCint}.

While the preceding discussion made use of $dP_9$ geometry to describe the nature of the singularity, the prescriptions for distinguishing roots and computing $U(1)$ charges relied only on local intersection data near $S_{\rm GUT}$.  To be sure, even though the object ${\cal{C}}$ was defined globally, its intersections with the roots $\{C_i\}$ and $\{C_a\}$ all occur in the neighborhood of $S_{\rm GUT}$.  In that sense, the collection of exceptional lines provides us with the construction of a noncompact divisor in the local geometry near $S_{\rm GUT}$ that yields a $U(1)$ symmetry capable of distinguishing the $\mathbf{10}$'s on different components of $\Sigma_{\mathbf{10}}${\footnote{Factoring of this noncompact divisor is of course trivially equivalent to factoring of the Higgs bundle spectral cover.}}.  Engineering an honest $U(1)$ in a compact model just requires us to provide a global extension of this noncompact divisor in a more general setting without the global structure of a $dP_9$ or K3.


\subsection{The ``Tate divisor"}
\label{subsec:tatediv}

Given a ``Tate model" of the form \eqref{tateform}, we propose that a suitable object that extends the noncompact divisor of our local model is the ``Tate divisor" defined by
\begin{equation}{\cal{C}}_{\rm Tate}:\quad a_0z^5+a_2z^3x+a_3z^2y+a_4zx^2+a_5xy \,,
\label{tatediv}\end{equation}
where it is understood that we take a suitable proper transform when passing to the resolution of $Y_4${\footnote{This is pretty important since ${\cal{C}}_{\rm Tate}$ is singular along the surface of $SU(5)_{\rm GUT}$ singularities where all 5 sheets come together.  Resolution of the $SU(5)_{\rm GUT}$ singularities separates the sheets and removes this singularity.}}.
In the neighborhood of $S_{\rm GUT}$ this behaves in precisely the same way as the object ${\cal{C}}$ above so that we can directly apply the local intersection data \eqref{localint} without having to work through the explicit resolution of singularities for each $Y_4$.  When this object becomes reducible into components ${\cal{C}}^{(m)}$ and ${\cal{C}}^{(n)}$ with $m$ and $n$ sheets, respectively, it reflects the fact that some of the $\mathbf{10}$ roots that degenerate along curves in $S_{\rm GUT}$ are distinguished from others and also provides us with a $U(1)$ that can make the distinction.  Explicitly, the $U(1)$ that does the job is a traceless combination
\begin{equation}\omega = n{\cal{C}}^{(m)}-m{\cal{C}}^{(n)} - \pi^*\delta \,,
\end{equation}
where $\delta$ is chosen to ensure that $\omega$ is orthogonal to the horizontal and vertical divisors of $Y_4${\footnote{We have to take a traceless combination because something with pure trace, like say ${\cal{C}}^{(m)}$, will have nontrivial intersection with the fiber class by virtue of the fact that $\ell_i\cdot F = 1$ \eqref{elldefs}.}}.

The object ${\cal{C}}_{\rm Tate}$ is an example of a ``spectral divisor", an object that we introduced in \cite{Marsano:2010ix} for defining global $G$-fluxes in a way that exploited the connection to Heterotic using essentially the same type of reasoning as above.  What makes ${\cal{C}}_{\rm Tate}$ special is that there is a simple way to take a local splitting of ${\cal{C}}_{\rm Tate}$, by which we mean a situation in which ${\cal{C}}_{\rm Tate}$ appears to contain several distinct components when restricted to the local neighborhood of $\pi^*S_{\rm GUT}$, and extend it into a global splitting that honestly divides the full ${\cal{C}}_{\rm Tate}$ into components.  The reason for this is that the meromorphic section $t$ defined as
\begin{equation}t = \frac{y}{x}\end{equation}
is actually holomorphic when restricted to ${\cal{C}}_{\rm Tate}$ so that we can write the defining equation \eqref{tatediv} as{\footnote{We use the fact that $y=t^3$ and $x=t^2$ on ${\cal{C}}_{\rm Tate}$.}}
\begin{equation}a_0z^5+a_2z^3t^2+a_3z^2t^3+a_4zt^4+a_5t^5 \,.
\end{equation}
As a homogeneous polynomial of degree 5 in $z$ and $t$ it is fairly easy to choose the $a_m$'s so that this object splits.  The algebra is completely equivalent to the splitting of Higgs bundle spectral covers in local model building \cite{Marsano:2009gv,Blumenhagen:2009yv,Marsano:2009wr,Grimm:2009yu}.

\subsection{$G$-Fluxes, Matter Surfaces, and Chiral Spectrum}
\label{subsec:specdiv}

The ``Tate divisor" can also be used to construct $G$-fluxes for chirality as described in \cite{Marsano:2010ix}.  We do not repeat that discussion here but simply remind the reader of the prescription to describe these $G$-fluxes as $(1,1)$-forms inside ${\cal{C}}_{\rm Tate}$.  Such a $(1,1)$-form effectively defines a $(2,2)$-form in the fourfold obtained by resolving the singularities of $Y_4$ that, as an object inside ${\cal{C}}_{\rm Tate}$, is sensitive to the details of that resolution.  It is this last fact that allows $G$ to know about the degenerate roots and integrate nontrivially over the matter surfaces, which we define as the surface in $\tilde{Y}_4$ that maps to a curve of singularities in $Y_4$ under the blow-down map.  In general, these surfaces should have the structure of a resolved root fibered over a curve. 

The $G$-fluxes needed to engineer chiral matter can be defined when ${\cal{C}}_{\rm Tate}$ is irreducible and, in fact, we can even use a different ``spectral divisor" to construct $G$-flux whose local defining equation near the surface of $SU(5)_{\rm GUT}$ singularities is equivalent to that of ${\cal{C}}_{\rm Tate}$ despite differing globally.  Nevertheless, to keep things simple we will always use ${\cal{C}}_{\rm Tate}$ to study $G$-fluxes in this paper.  In that case, we get a particularly nice type of $G$-flux when ${\cal{C}}_{\rm Tate}$ is reducible and can be used to construct $U(1)$s.  We simply take a $(1,1)$-form $\rho$ in $B_3$, pull it back to each component of ${\cal{C}}_{\rm Tate}$, and take a traceless combination of the two
\begin{equation}G = \left(n{\cal{C}}^{(m)}-m{\cal{C}}^{(n)}\right)\cdot \pi^*\rho - G_0 \,,
\label{U1flux}\end{equation}
where $G_0$ is a subtraction term that we must include to ensure that $G$ is orthogonal to horizontal and vertical divisors in $Y_4$.  The form \eqref{U1flux} is very much in the spirit of how $G$-fluxes in F-theory should heuristically be related to $U(1)$ flux as
\begin{equation}G\sim \omega_i\wedge F_i\end{equation}
with $\omega_i$ the (1,1)-form specifying a $U(1)$ and $F_i$ the flux associated to that $U(1)$.



Let us turn now to determining the chiral spectrum, which involves integrating $G$ over various matter surfaces.  In \cite{Marsano:2010ix} we studied ways to do this with a focus on rephrasing the computation as one that can be done in terms of divisor classes and intersections in $Y_4$.  Rather than executing gymnastics like that, we just review here how the Higgs bundle picture emerges.

The $G$-flux that we build will have the generic form
\begin{equation}G = {\cal{G}} - G_0\,,
\label{Gdef}\end{equation}
where ${\cal{G}}$ is obtained from the ``Tate divisor" (as a holomorphic surface inside ${\cal{C}}_{\rm Tate}$) and $G_0$ is some holomorphic surface in $Y_4$ whose contribution we must add to ensure that $G$ is orthogonal to horizontal and vertical divisors in $Y_4$.
Chiral matter localizes on curves of singularities $\Sigma$ where 2-cycles $C$ degenerate and the flux that they couple to is obtained by restricting $G$ to the matter surface 
`$\Sigma\times C$'{\footnote{We put $\Sigma\times C$ in quotes because the surface does not have to be a direct product.}} and integrating it over $C$.  The net chirality follows from integrating $G$ over the entire matter surface and this is the computation we ultimately want to describe.

Intersecting $G$ with any holomorphic surface will require a special type of calculation.  From the form \eqref{Gdef} we see that intersections of ${\cal{G}}$ with generic holomorphic surfaces will generally be cancelled by intersections with $G_0$.  The types of intersections for which this does not happen are those that occur at the singular locus, to which ${\cal{G}}$ is sensitive but $G_0$ is not.  By construction, then, $G$ only has nonzero integral over surfaces that include degenerate cycles so for any computation we can focus our attention on the neighborhood of the curve of singularities in question.  For $SU(5)_{\rm GUT}$-charged matter, it is enough to focus on the limiting behavior of ${\cal{C}}_{\rm Tate}$ and ${\cal{G}}$ near the surface of $SU(5)_{\rm GUT}$ singularities, which we do by taking $y/x=t\rightarrow 0$ and $z\rightarrow 0$ with $s=z/t$ fixed.  Recall that ${\cal{C}}_{\rm Tate}$ is described by
\begin{equation}a_0z^5+a_2z^3x+a_3z^2y+a_4zx^2+a_5xy \,.
\end{equation}
In the limit, then, ${\cal{C}}_{\rm Tate}$ becomes
\begin{equation}{\cal{C}}_{\rm Tate}\rightarrow t^5\left(b_0s^5+b_2s^3+b_3s^2+b_4s+b_5\right)\label{tatehiggs}\end{equation}
with the $b_m$'s now restricted to sections on $S_{\rm GUT}$ and $s$ a section of the canonical bundle on $S_{\rm GUT}$.  The term in $(\,)$'s here is nothing other than the Higgs bundle spectral cover, ${\cal{C}}_{\text{Higgs}}$ \cite{Donagi:2009ra}.  The limiting behavior of ${\cal{G}}$ also produces something familiar; it gives us a divisor $\gamma$ inside ${\cal{C}}_{\text{Higgs}}$ that we would like to interpret as the corresponding object in the local model.


It is interesting to see in some detail how the approach of ${\cal{C}}_{\rm Tate}$ to the surface of $SU(5)_{\rm GUT}$ singularities takes place.  For starters, the $t^5$ factor tells us that every (local) sheet of ${\cal{C}}_{\rm Tate}$ meets the surface of $SU(5)_{\rm GUT}$ singularities once above each point on $S_{\rm GUT}$.  This reflects the fact that each exceptional line $\ell_i$ in the $dP_9$ picture transforms in a $\mathbf{10}$ of $SU(5)_{\rm GUT}$ and must therefore intersect some of the $SU(5)_{\rm GUT}$ roots at each point on the surface of singularities.  This accounts for the `1' in the local intersection data \eqref{localint} and has no effect on any of our computations when the $G$-flux is chosen to be traceless{\footnote{Indeed, from this point of view, the traceless condition of the $G$-flux that we imposed in \cite{Marsano:2010ix} is really just the statement that we require our $G$-flux to be orthogonal to all $SU(5)_{\rm GUT}$ roots.  Trace parts may have an important role to play, however.  It has been known for some time \cite{Donagi:2008kj} that what is commonly referred to as hypercharge flux must include some component from the bulk in the sense that it is not a flux in the pure $U(1)_Y$ direction.  A $G$-flux constructed within the ``Tate divisor" that has a pure trace piece is exactly the sort of thing we need to give a proper definition of this hypercharge flux.}}.

The rest of the approach is captured by the behavior of the Higgs bundle spectral cover.  Consider first the curve of $SO(10)$ singularities, $\Sigma_{10}$, which sits at $y=b_5=0$.  Within the second piece of \eqref{tatehiggs} there is a distinguished curve, $t=b_5=0$, that coincides with $\Sigma_{10}$.  It captures part of the ``Tate divisor" that lands exactly on $\Sigma_{10}$ as we approach the surface of $SU(5)_{\rm GUT}$ singularities.  The curve $t=b_5=0$ is familiar from local model building: it is the ``local $\mathbf{10}$ matter curve" by which we mean the $\mathbf{10}$ matter curve of the local model that sits inside the Higgs bundle spectral cover ${\cal{C}}_{\text{Higgs}}$.  Here we call this curve $\Sigma_{10,\text{Higgs}}$ and it appears naturally in the Calabi-Yau fourfold as describing the limiting behavior of the specific (local) sheet of ${\cal{C}}_{\rm Tate}$ that lands directly on the curve of $SO(10)$ singularities.  This is just another way of saying that $\Sigma_{10,\text{Higgs}}$ describes a cross section of the exceptional line $\ell_i$ that is dual to the root $C_i$ whose degeneration causes the $SO(10)$ enhancement.

From this point of view, it is clear that restricting $G$ to the matter surface $`\Sigma_{\mathbf{10}}\times C$' is equivalent to projecting ${\cal{G}}$ onto $\Sigma_{10,\text{Higgs}}$, which is isomorphic to the $\mathbf{10}$ matter curve in the copy of $S_{\rm GUT}$ that sits in the section{\footnote{Note that the singularities of $Y_4$, including the surface of singularities above $S_{\rm GUT}$, do not lie in the section.}}.
This reproduces the standard result from Higgs bundles that charged fields are sections of
\begin{equation}H^m(\Sigma_{10},K_{\Sigma_{10}}^{1/2}\otimes i_*\gamma)\qquad m=0,1
\end{equation}
with $\Sigma_{10}$ the curve of $SO(10)$ enhancements in $S_{\rm GUT}$. 
  Recall that $\gamma$ is a divisor in $\mathcal{C}_{\rm Higgs}$ that represents the limiting behavior of $\mathcal{G}$ in our global approach and specifies the ``local flux" in a local model.  
The map $i$ is just the usual embedding  $i:\Sigma_{10,\text{Higgs}}\rightarrow \Sigma_{10}$.

There is a similar story for the $\mathbf{\overline{5}}$ curve.  
To see this, recall that the surface of $SU(5)_{\rm GUT}$ singularities is concretely given by the intersection of the ``Tate form" \eqref{tateform}{\footnote{This corresponds to $\hat{y}=z=0$ in the corresponding Weierstrass model.  The equation for $y$ reflects the shift that is needed to go from Tate to Weierstrass form.}}
\begin{equation}y-\frac{1}{2}(b_3x+b_5z^2) = z = 0 \,.
\end{equation}
With our current variables, the first of these corresponds to{\footnote{This relation is just setting the $Y$ coordinate of the Weierstrass form to zero.}}
\begin{equation}t^3-\frac{z^2}{2}(b_3s^2+b_5)=0 \,.
\end{equation}
Now, the $\mathbf{\overline{5}}$ curve $\Sigma_{\overline{5},\text{Higgs}}$ of the local model with Higgs bundle spectral cover ${\cal{C}}_{\text{Higgs}}$ is described by
\begin{equation}b_3s^2+b_5= b_0s^4+b_2s^2+b_4=0 \,,
\end{equation}
which we can essentially think of as the intersection of $b_3s^2+b_5=0$ with ${\cal{C}}_{\text{Higgs}}$.  As we send $t\rightarrow 0$, then, the curve $\Sigma_{\overline{5},\text{Higgs}}$ lands directly on the curve of $SU(6)$ singularities.

As in the case of $\mathbf{10}$'s, this suggests that restricting $G$ to the matter surface $`\Sigma_{\overline{5}}\times C$' is equivalent to projecting ${\cal{G}}$ onto the curve of $SU(6)$ singularities.  This projection is now 2-1 for the usual reasons and leads to the familiar result from Higgs bundles that $\mathbf{\overline{5}}$'s and $\mathbf{5}$'s are sections of
\begin{equation}H^m(\Sigma_{\overline{5}},K_{\Sigma_{\overline{5}}}^{1/2}\otimes \nu_*\gamma)\qquad m=0,1 \,,
\end{equation}
where $\nu$ is the 2-1 covering map from the curve $\Sigma_{\overline{5},\text{Higgs}}$ in ${\cal{C}}_{\text{Higgs}}$ to the curve $\Sigma_{\overline{5}}$ of $SU(6)$ singularities inside $S_{\rm GUT}$.  We know from the seminal work of \cite{Hayashi:2008ba} that we should actually take $\Sigma_{\overline{5}}$ to be the normalization of the curve of $SU(6)$ singularities.  We understand this statement here in the following way.  Singularities of $\Sigma_{\overline{5}}$ come from nodes where the singularity type enhances.  The two branches of $\Sigma_{\overline{5}}$ that meet at such a node are places where homologously distinct cycles are degenerating so those branches are separated in the lift to ${\cal{C}}_{\text{Higgs}}$ (which we view as the limiting behavior of ${\cal{C}}_{\rm Tate}$) upon resolution.

Note that this picture makes clear why matter fields are associated with sections on matter curves inside the Higgs bundle spectral cover.  Those curves are capturing the actual curves of singularities, as seen by the $G$-flux, in the fourfold.  It is along the curves of singularities, not their projection to the section (which does not meet the singularities), that the wrapped branes are found.

Even though we can say something about the actual cohomology groups associated to various charged matter fields, let us say a few more things about the simpler question of net chirality.  On general grounds we expect that the net chirality on a matter curve $\Sigma$ is obtained by integrating $G$ over the corresponding matter surface `$\Sigma\times C$' and, from our discussion above, this amounts to integrating $\gamma$ over the ``local matter curves" $\Sigma_{R,\text{Higgs}}$ inside the Higgs bundle spectral cover.  The only reason for the appearance of ${\cal{C}}_{\text{Higgs}}$, though, was our insistence on studying the chiral spectrum on matter curves that sit inside the surface of $SU(5)_{\rm GUT}$ singularities{\footnote{The presence of the ambient surface of singularities was also responsible for the gymnastics associated with imposing tracelessness and dropping the $t^5$}}.  We can give a description of the computation of chiral matter that does not make reference to ${\cal{C}}_{\text{Higgs}}$ as follows.  First, we embed the curve of singularities $\Sigma$ associated to a matter surface `$\Sigma\times C$' into ${\cal{C}}_{\rm Tate}$.  Then, we determine the intersection of ${\cal{G}}$ with that surface.

The nice thing about this prescription is that it works equally well for matter curves that do not sit inside the surface of $SU(5)_{\rm GUT}$ singularities.  In principle, then, we should be able to determine the net chirality of $SU(5)_{\rm GUT}$-singlet fields that carry $U(1)$ charge in $SU(5)_{\rm GUT}$ models that engineer an extra $U(1)$.  We hope to report soon on models in which this computation is explicitly carried out.


\subsection{D3-brane tadpole}

Let us now make some remarks about the D3-brane tadpole.
To compute the D3-brane tadpole induced by a $G$-flux, what we need to evaluate is a self-intersection
\begin{equation}\int_{Y_4}G\wedge G \,,
\label{Gsquared}\end{equation}
which we can think of in two pieces
\begin{equation}{\cal{G}}\cdot ({\cal{G}}-G_0) - G_0\cdot ({\cal{G}}-G_0)\end{equation}
with ${\cal{G}}$ and $G_0$ defined as in \eqref{Gdef}.  The second term here vanishes by construction because $G_0$ is an honest surface in $Y_4$ that is constructed as a linear combination of intersections of horizontal and vertical divisors.  What we are after, then, is the first term.  We can think of this as integrating $G$ over the surface $S_{\cal{G}}$ in the resolution of $Y_4$ that is determined by ${\cal{G}}$.  As usual, any integral of $G$ over a surface will receive contributions only from places where ${\cal{G}}$ meets singularities that have to be resolved.  So, what we want is to look at the full locus where ${\cal{G}}$ meets singularities and restrict ${\cal{G}}$ to that locus.  If we consider, for instance, the surface of $SU(5)_{\rm GUT}$ singularities then the restriction of ${\cal{G}}$ is the curve $\gamma$ that sits inside the Higgs bundle spectral cover ${\cal{C}}_{\text{Higgs}}$.  From the surface of $SU(5)_{\rm GUT}$ singularities, then, we get a contribution to \eqref{Gsquared} from each point in the self-intersection $\gamma\cdot_{ {\cal{C}}_{\text{Higgs}}}\gamma$.  The setup is in fact almost identical to the computation of chiral matter except that we have replaced the matter curve by the curve $\gamma$ in order to reflect the fact that we are integrating over the curve $S_{\cal{G}}$ determined by ${\cal{G}}$.  When we do this, though, we must be a little careful because ${\cal{G}}$ is defined on a union of $\ell_i\sim (x_9-e_9)-C_i$'s and the contribution to \eqref{Gsquared} that comes from a singular point arises from the $C_i$.  As the sign is opposite from what we have when we integrate $G$ over a matter surface we find that the contribution to \eqref{Gsquared} is in fact $-\gamma\cdot_{ {\cal{C}}_{\text{Higgs}}}\gamma$.  Note that this type of reasoning is not at all new and should be familiar from studies of Heterotic/F-theory duality \cite{Friedman:1997yq,Curio:1998bva,Donagi:2008ca,Donagi:2008kj,Hayashi:2009ge}.

More generally, we expect \eqref{Gsquared} to be computed by the restriction of ${\cal{G}}\cdot_{{\cal{C}}_{\rm Tate}}{\cal{G}}$ to the locus of singularities with the appropriate signs.  Note that ${\cal{G}}\cdot_{ {\cal{C}}_{\rm Tate}} {\cal{G}}$ is a curve inside ${\cal{C}}_{\rm Tate}$ so it will generically miss curves of singularities or isolated point singularities.  This means that \eqref{Gsquared} should get contributions only from surfaces of singularities in accord with a conjecture of \cite{Grimm:2009yu}.  The discussion for a surface of $SU(5)_{\rm GUT}$ singularities should generalize to something like the following for the D3-brane tadpole induced by $G$-flux when there are multiple surfaces of singularities that do not intersect one another
\begin{equation}Q_{D3,\text{induced}}=-\frac{1}{2}\int_{Y_4}G\wedge G = \frac{1}{2}\sum_{\text{surfaces of singularities,}i}\int_{ {\cal{C}}_i}\gamma_i^2 \,,
\label{D3tadpole}\end{equation}
where ${\cal{C}}_i$ is essentially the Higgs bundle spectral cover for the $i$th surface and $\gamma_i$ the limit of ${\cal{G}}$.  It is actually not hard to see that a spectral cover for an $SU(n)$ Higgs bundle emerges from the limiting behavior of ${\cal{C}}_{\rm Tate}$ near any surface of singularities whose commutant inside $E_8$ is an $SU(n)$ group or a product of such groups.  Looking at the behavior of ${\cal{C}}_{\rm Tate}$ for more general types of singularities and interpreting the result would be very interesting but is beyond the scope of this paper.


\subsection{Remarks on Flux Quantization}

We now turn to flux quantization, an issue that we somewhat neglected in \cite{Marsano:2010ix}.  Rather than simply rewriting Witten's famous condition \cite{Witten:1996md}, let us recall one of the original observations that points to the need for modifying the $G$-flux quantization rule.  This will motivate a consistency check that we perform later.

The D3-brane tadpole in Calabi-Yau fourfold compactifications of F-theory receives both a flux contribution, which we have described, and a geometric contribution from the Euler character
\begin{equation}n_{D3} = \frac{\chi}{24}-\frac{1}{2}G^2 \,.
\end{equation}
For smooth Calabi-Yau fourfolds, it was shown in \cite{Sethi:1996es} that $\chi$ is always divisible by 6 but not necessarily by 12.  When $\chi/24$ fails to be an integer, the quantization of $G$ must be modified in order to account for this fact and ensure an integral induced D3-brane charge.  In \cite{Sethi:1996es}, it is proven that $\chi$ is always divisible by 24 when the fourfold admits a smooth Weierstrass description. More recent work \cite{Marsano:2009ym,Collinucci:2010gz} uses similar reasoning to argue directly that $G$-fluxes are integrally quantized in such cases {\footnote{\cite{Marsano:2009ym} showed that $c_2(\tilde{Y}_4)$ is even whenever the class $c_1(B_3)^2-c_2(B_3)$ is even in $B_3$ and argued that compactifications in which $c_1(B_3)^2-c_2(B_3)$ is odd are necessarily Lorentz-violating.  The work \cite{Collinucci:2010gz} actually proved that $c_1(B_3)^2-c_2(B_3)$ is always even.}}.  When our $Y_4$ is singular, though, we need to ask questions about $\chi$ and $G$-flux quantization on the smooth resolution $\tilde{Y}_4$ which is not of this type.  We can think of $\tilde{Y}_4$ as a smooth Calabi-Yau fourfold but cannot use any special properties attributed to fourfolds that can be realized as smooth Weierstrass models.

The connection between $\chi$ and $G$-flux quantization arises from a few simple facts about Calabi-Yau fourfolds.  First of all, the Todd genus of a smooth Calabi-Yau fourfold $\tilde{Y}_4$ is 2, which implies that
\begin{equation}\lambda^2 = 480+\frac{\chi}{3} \,,
\label{c2chi}\end{equation}
where  $\lambda$ denotes the second Chern class
\begin{equation}\lambda = c_2(\tilde{Y}_4) \,.
\end{equation}
Using this, we can write the induced D3-brane charge as
\begin{equation}\begin{split}n_{D3} &= 60 - \frac{1}{2}\left[G^2-\left(\frac{\lambda}{2}\right)^2\right] \\
&= 60 - \frac{1}{2}\left[\left(\alpha - \frac{\lambda}{2}\right)^2 - \left(\frac{\lambda}{2}\right)^2\right] \\
&= 60 - \frac{1}{2}\left[\alpha^2-\alpha\cdot \lambda\right] \,,
\end{split}\end{equation}
where we implicitly defined
\begin{equation}\alpha = G+\frac{\lambda}{2} \,.
\end{equation}
Writing things in this way is helpful because the intersection of any holomorphic surface $\tilde{S}$ inside $\tilde{Y}_4$ with the class $\lambda$ satisfies \cite{Witten:1996md}
\begin{equation}\tilde{S}\cdot \lambda = \tilde{S}^2\text{ mod }2\end{equation}
so that the induced D3-brane charge is guaranteed to be an integer whenever $\alpha$ is an integral class
\begin{equation}\alpha = G+\frac{\lambda}{2} \in H^4(\tilde{Y}_4,\mathbb{Z}) \,.
\end{equation}
This is just the quantization law derived by Witten \cite{Witten:1996md}.  We emphasize here that the joint geometric and flux contributions to the D3-brane tadpole reflect a connection between $\chi$ and the $G$-flux quantization law that is embodied by \eqref{c2chi}.  This will allow a useful consistency check on any proposed quantization rule.

\subsubsection{New surfaces in $\tilde{Y}_4$}
\label{subsubsec:newsurfaces}

We now turn to the quantization of $G$-flux directly, which is to say a characterization of the odd part of $\lambda$.  Because $\lambda$ is always even in smooth Weierstrass models, such as the fourfold we get by deforming away the singularities of $Y_4$, we expect that any odd piece of $c_2(\tilde{Y}_4)$ comes from holomorphic surfaces in $\tilde{Y}_4$ that do not survive the blow-down map $\tilde{Y}_4\rightarrow Y_4$.  Surfaces of this type include matter surfaces, which have the form $`C\times \Sigma$' for some degenerating root $C$, as well as the surfaces (or rather formal linear combinations of surfaces) that can be described in the spectral divisor formalism \cite{Marsano:2010ix}{\footnote{We do not claim to have identified a distinguished linearly independent basis for the set of surfaces under consideration.}}.  The latter are of the form ${\cal{S}}-S_0$ for ${\cal{S}}$ a divisor inside ${\cal{C}}_{\rm Tate}$ and $S_0$ the usual subtraction piece that ensures ${\cal{S}}-S_0$ is trivial after the blow-down map.  Given $\tilde{S}$, we will use the notation $\tilde{S}_{\cal{S}}$ to denote a surface class in $\tilde{Y}_4$ that is constructed via the ${\cal{S}}-S_0$ procedure from a divisor ${\cal{S}}$ in ${\cal{C}}_{\rm Tate}$.  

We will later be interested in computing intersections of these new surfaces $\tilde{S}_{\cal{S}}\cdot_{\tilde{Y}_4}\tilde{S}_{\cal{S}}$.  In principle, we can use our knowledge of the local geometry near singularities of $Y_4$ to do this.  For simplicity, let us suppose that our $Y_4$ has a single surface of $SU(5)_{\rm GUT}$ singularities and no others.  In that case, the local intersection data \eqref{localint} from the $dP_9$ picture tells us that the computation should reduce to one within the Higgs bundle spectral cover.  Suppose that a divisor ${\cal{S}}$ in ${\cal{C}}_{\rm Tate}$ restricts to a divisor $s$ inside ${\cal{C}}_{\text{Higgs}}$.  In the same way that we computed the D3-brane tadpole contribution \eqref{D3tadpole}, we can use the local intersection data \eqref{localint} to obtain
\begin{equation}\tilde{S}_{{\cal{S}}_1}\cdot_{\tilde{Y}_4} \tilde{S}_{{\cal{S}}_2} = -s_1\cdot_{\cal{C}_{\text{Higgs}}}s_2 - (p_*s_1)\cdot_{S_{\rm GUT}}(p_*s_2) \,,
\end{equation}
where $p$ is the projection map
\begin{equation}
p:{\cal{C}}_{\text{Higgs}}\rightarrow S_{\rm GUT} \,.
\end{equation}
The first term is familiar from the discussion preceding \eqref{D3tadpole}.  The new term involving $p_*s_i$'s accounts for the fact that ${\cal{S}}$ need not be traceless.  It effectively computes the contribution from the `1' in \eqref{localint}.

We can tabulate a few interesting results here.  When ${\cal{C}}_{\text{Higgs}}$ is generic, which means among other things that it \emph{does not split}, the divisors of ${\cal{C}}_{\text{Higgs}}$ are of the form
\begin{equation}\sigma\cdot_X {\cal{C}}_{\text{Higgs}}\qquad\text{and}\qquad p^*\Sigma \,,
\end{equation}
where $\Sigma$ is a curve in $S_{\rm GUT}$, $X=\mathbb{P}({\cal{O}}\oplus K_{S_{\rm GUT}})$ is the usual ambient space in which ${\cal{C}}_{\text{Higgs}}$ is embedded for ease of study, and $\sigma$ is a section of the $\mathbb{P}^1$-fibration $X$ that transforms trivially as we move along $S_{\rm GUT}${\footnote{This is the conventional notation introduced to the F-theory literature by \cite{Donagi:2009ra}.}}.  Given this, we know that any divisor ${\cal{S}}$ must restrict to a combination
\begin{equation}
{\cal{S}}\rightarrow n\left(\sigma\cdot_X {\cal{C}}_{\text{Higgs}}\right)+p^*\Sigma \,.
\end{equation}
This allows us to give general formulas.  Writing $c_1$ as shorthand for $c_1(S_{\rm GUT})$ as usual and denoting the class of ${\cal{C}}_{\text{Higgs}}$ inside $X$ as
\begin{equation}{\cal{C}}_{\text{Higgs}}=5\sigma + \pi_X^*\eta\,,\qquad \pi_X:X\rightarrow S_{\rm GUT}\end{equation}
we have that
\begin{equation}\begin{split}\left(\tilde{S}_{\sigma\cdot {\cal{C}}_{\text{Higgs}}}\right)^2 &= -30c_1^2+11c_1\eta - \eta^2 \\
\left(\tilde{S}_{\sigma\cdot {\cal{C}}_{\text{Higgs}}}\right)\cdot_{\tilde{Y}_4}\left(\tilde{S}_{p^*\Sigma}\right) &= -6\Sigma\cdot_{S_{\rm GUT}}(\eta-5c_1) \\
\left(\tilde{S}_{p^*\Sigma_1}\right)\cdot_{\tilde{Y}_4}\left(\tilde{S}_{p^*\Sigma_2}\right) &= -30\Sigma_1\cdot_{S_{\rm GUT}} \Sigma_2\,.
\end{split}\end{equation}
We have abused notation somewhat here.  By $\tilde{S}_{\sigma\cdot_X {\cal{C}}_{\text{Higgs}}}$ we mean $\tilde{S}_{{\cal{S}}}$ for a divisor ${\cal{S}}$ that restricts to $\sigma\cdot_X {\cal{C}}_{\text{Higgs}}$ in the Higgs bundle spectral cover.

\subsubsection{$G$-flux quantization}

A rule for $G$-flux quantization is naturally motivated from the study of Higgs bundles.  In the latter setting, the object $\gamma$ that represents the restriction of ${\cal{G}}$ to ${\cal{C}}_{\text{Higgs}}$ satisfies the rule
\begin{equation}
\gamma + \frac{r}{2}\in H^2({\cal{C}}_{\text{Higgs}},\mathbb{Z}) \,,
\end{equation}
where $r$ is the ramification divisor of the covering
\begin{equation}p:{\cal{C}}_{\text{Higgs}}\rightarrow S_{\rm GUT} \,.
\end{equation}
The divisor $r$ naturally descends from a divisor class $\tilde{r}$ inside ${\cal{C}}_{\rm Tate}$ obtained by taking the ramification divisor of the covering ${\cal{C}}_{\rm Tate}\rightarrow B_3$ and removing the component along the surface of $SU(5)_{\rm GUT}$ singularities. 
The proposed quantization rule is now that ${\cal{G}}$ should be chosen so that
\begin{equation}{\cal{G}}+\frac{\tilde{r}}{2}\in H^4({\cal{C}}_{\rm Tate},\mathbb{Z}) \,.
\end{equation}
This is equivalent to the claim that the combination $\tilde{r}-r_0$ captures the odd part of $c_2(\tilde{Y}_4)$ where $r_0$ represents our usual subtraction term.  In equations, the conjecture is
\begin{equation}\frac{1}{2}\left[\lambda(\tilde{Y}_4)-(\tilde{r}-r_0)\right]\in H^4(\tilde{Y}_4,\mathbb{Z}) \,.
\end{equation}

As usual, this is something that has been conjectured, at least in some form, in the setting of Heterotic/F-theory duality.  In addition to describing everything intrinsically in terms of F-theory and the geometry of $\tilde{Y}_4$, we would like to perform a consistency check that we do not believe has been studied previously using duality or otherwise.  This check has to do with the connection between $\lambda$ and the Euler character \eqref{c2chi}, which tells us that the odd part of the former is related to the part of the latter that fails to be divisible by 24.  More specifically, suppose we compute $\chi$ and $\lambda$ in a smooth deformation of $Y_4$ and follow the resulting classes as we move back to the singular $Y_4$.  The result of this is the naive $\chi$ and $\lambda$ that we would obtain by blindly applying formulae for smooth Weierstrass models without properly accounting for the singularity structure.  If we pull these back to the resolved geometry $\tilde{Y}_4$, they will differ from the actual $\chi(\tilde{Y}_4)$ and $\lambda(\tilde{Y}_4)$ there by shifts
\begin{equation}\delta\chi = \chi(\tilde{Y}_4) - \chi(Y_4)\,,\qquad \delta\lambda = \lambda(\tilde{Y}_4)-\lambda(Y_4)\,.\end{equation}
The general relation \eqref{c2chi} tells us that{\footnote{We use the fact that $\lambda(Y_{4})$ is an even class.}}
\begin{equation}\frac{\delta \chi}{12}+ \frac{1}{4}(\delta \lambda)^2 \in \mathbb{Z} \,.
\end{equation}
The odd part of $\delta\lambda$ is thus directly connected to the failure of $\chi$ to be divisible by 12.  In our $SU(5)_{\rm GUT}$ example, we conjecture that the odd part of $\delta\lambda$ is given by $(\tilde{r}-r_0)$ so the non-integer part of $(\delta\lambda)^2/4$ can be computed using the rules of section \ref{subsubsec:newsurfaces}
\begin{equation}\tilde{S}_{(\tilde{r}-r_0)}\cdot_{\tilde{Y}_4}\tilde{S}_{(\tilde{r}-r_0)} = -15(32c_1^2-25\eta c_1+5\eta^2)\,.
\end{equation}
We can now compare this to the formulae conjectured in \cite{Blumenhagen:2009yv} for the shift in Euler character due to the $SU(5)_{\rm GUT}$ singularity.  This leads to the result
\begin{equation}\delta\chi = -15(488c_1^2-211\eta\cdot c_1 + 23\eta^2)\,.
\end{equation}
From this, it is easy to verify that
\begin{equation}\frac{\delta\chi}{12} + \frac{1}{4}\left(\delta\lambda\right)^2\in \mathbb{Z}\,.\end{equation}
We believe this represents nice evidence in favor of the conjectures of \cite{Blumenhagen:2009yv} relating to the Euler character and the conjectured odd part of $\lambda$.

Note that the conjecture for $\delta\lambda$ will take a slightly different form when $U(1)$s are engineered.  This is because a split ``Tate divisor" ${\cal{C}}_{\rm Tate}$ will lead to a split Higgs bundle and what enters the quantization condition for $\gamma$ in that case are the ramification divisors $r_n$ of the different components of ${\cal{C}}_{\text{Higgs}}$.  We therefore expect that the shift in Euler character $\delta\chi$ will be similarly affected.

The above story also provides more evidence that surfaces of singularities (and likely the local behavior of the ``Tate divisor") play a dominant role in determining the Euler character.  If we have multiple non-intersecting surfaces of singularities in $Y_4$, we expect a local Higgs bundle description (or something like it) near each one and each will carry with it a new potentially odd contribution to $\lambda$.  Shifts in $\lambda$ go hand in hand, according to \eqref{c2chi}, with shifts in the Euler character.


\section{M5 instantons, $G$-flux, and charged couplings}
\label{sec:M5}

We now move to a simple application of our description of $G$-flux: the influence of flux on M5 instantons.  More generally, we would like to know, for a given choice of Calabi-Yau fourfold and $G$-flux, which divisors can support M5-instantons that generate nonzero couplings and whether any of those couplings involve charged fields, by which we mean states that arise from M2-branes that wrap degenerate cycles.

For starters, recall that in F-theory we are really studying backgrounds of the type $\mathbb{R}^{3,1}\times B_3$ in IIB string theory with axio-dilaton that varies over $B_3$.  Nonperturbative superpotential corrections can arise not just from D3-instantons \cite{Witten:1996bn,Ganor:1996pe,Denef:2004dm,Saulina:2005ve,Kallosh:2005gs,Denef:2005mm}  wrapped on divisors in $B_3$ but also from gaugino condensation \cite{Tripathy:2002qw,Gorlich:2004qm,Denef:2004dm}  which may occur along divisors in $B_3$ that support non-Abelian gauge groups.  For applications to F-theory GUT model-building the visible sector will not undergo gaugino condensation but, as we will review in the next section, hidden sector gauge groups can often appear and the corresponding gaugino condensates can play an important role in stabilizing K\"ahler moduli.

In this paper, we adopt the perspective of M-theory to describe D3-instantons as a limit of M5-instantons that are wrapped on vertical divisors of the elliptically fibered Calabi-Yau fourfold $Y_4$.  These vertical divisors are complex threefolds $D$ of the form $\pi^*S_2$ where $S_2$ is a holomorphic divisor in $B_3$.  In the F-theory limit, where the volume of the elliptic fiber of $Y_4$ is taken to zero, these M5-instantons descend to D3-instantons wrapped on $S_2$.  


In general, there are two reasons that an M5-instanton wrapping a vertical divisor fails to contribute a superpotential coupling $W_{inst}$

\begin{itemize}
\item $W_{inst}=0$ if there are not precisely two fermion zero modes on the M5-brane worldvolume

\item $W_{inst}$ may be zero if the restriction of the $G$-flux to the M5 worldvolume is nontrivial
\end{itemize}

Before proceeding, let us say a few more words about the second point.  The influence of $G$-flux is evident already from the fact that the complexified K\"ahler modulus{\footnote{$C_6$ is the magnetic dual of the 3-form gauge potential $C_3$ in M-theory.}} $T=\int_D(\text{vol}_D+iC_6)$ shifts under gauge transformations of the M-theory 3-form $C_3$ when the restriction of $G_4$ to the M5 is nontrivial.  To see why, recall that under $C_3\rightarrow C_3+d\Lambda_2$ the 6-form $C_6$ shifts as
\begin{equation}
C_6\rightarrow C_6+\frac{1}{2}\Lambda_2\wedge G_4 \,.
\end{equation}
This is a general consequence of the gauge invariance of $G_7=\ast G_4$ and the Bianchi identities
\begin{equation}
dG_7 = -\frac{1}{2}G_4\wedge G_4 \,,
\end{equation}
which implies that
\begin{equation}
G_7=dC_6 - \frac{1}{2}C_3\wedge G_4 \,.
\end{equation}
As a result, the complexified K\"ahler modulus $T=\int_D (\text{vol}_D+iC_6)$ and hence the classical action of the M5 shifts under gauge transformations of the M-theory 3-form $C_3\rightarrow C_3 + d\Lambda_2$ according to
\begin{equation}
\delta T = \frac{i}{2}\int_D \Lambda_2\wedge G_4 \,.
\end{equation}
An interesting special case of this arises when we have a $U(1)$ symmetry of the type considered in section \ref{subsec:dp9U1}, by which we mean a harmonic $(1,1)$-form $\omega$ on which we can reduce $C_3$ as $C_3=A_1\wedge \omega$.{\footnote{Here $A_1$ gives a 3-dimensional gauge field on $\mathbb{R}^{2,1}$ in M-theory that descends to a 4-dimensional gauge field on $\mathbb{R}^{3,1}$ in the F-theory limit when $\omega$ has exactly 1 leg on the elliptic fiber.}}  A gauge transformation with respect to this $U(1)$ corresponds to the special choice $\Lambda_2 = \phi\wedge \omega$ with $\phi$ the product of a constant 0-form on the Calabi-Yau fourfold and a nontrivial function of the Minkowski directions.  In that case, the classical M5-instanton action $e^{-T}$ transforms as though it carries charge
\begin{equation}\label{qM5}
q_{\text{M5}} = \frac{1}{2}\int_D\omega\wedge G_4 \,.\end{equation}
Any coupling generated by the M5-instanton must be gauge invariant under all potential gauge transformations of $C_3$.  When $G_4$ restricts trivially to the M5 worldvolume this is not a problem and, in that case, the fermi zero mode structure should represent the only obstacle to obtaining a nonzero coupling.  When $G_4$ has nontrivial restriction, we will argue that this can be accomplished in some cases by inserting wrapped M2-brane states{\footnote{What is really happening is that the flux induces a source term for the chiral 2-form $b_2$ so that the vacuum solutions about which we expand are solitons that describe suitable M2s ending on the M5.}}.  In those cases the M5 generates a coupling, again provided the fermi zero mode structure is appropriate.  When this is not possible, the M5 should not generate any coupling.

Note that invariance under bulk gauge transformations of $C_3$ is a stronger statement than invariance under $U(1)$ symmetries of the type described in section \ref{subsec:dp9U1}.  The full condition of bulk gauge invariance should incorporate the selection rules alluded to in \cite{Grimm:2011dj}{\footnote{As we understand it, gauge transformations corresponding to the massive $U(1)$s of \cite{Grimm:2011dj} correspond to choosing $\Lambda_2$ to be of the form $\phi\wedge\tilde{\omega}$ with $\phi$ as above and $\tilde{\omega}$ a 2-form on $Y_4$ that is not necessarily harmonic.}}.

In the rest of this section we discuss each of the two obstacles to obtaining a nonzero coupling in turn.  In section \ref{subsec:fermimodes}, we review the counting of fermion zero modes applying those results
to M5-instantons in the specific geometry of \cite{Marsano:2009ym} in section \ref{sec:application}.  Then in section \ref{subsec:M5Gflux} we discuss the effect of $G$-flux in a bit more detail, describe the impact of the $G$-fluxes that we discussed in section \ref{subsec:specdiv}, clarify how they could in principle generate charged couplings, and demonstrate that, for the $G$-fluxes described in section \ref{subsec:specdiv}, the restriction to M5 worldvolumes is generically trivial.  This means that vertical divisors with the right number of fermi zero modes should, in the absence of additional $G$-fluxes with nontrivial restriction, generate uncharged couplings that can play a role in the stabliization of K\"ahler moduli.

\subsection{Counting fermion zero modes}
\label{subsec:fermimodes}
\subsubsection{Review of fermions living on M5-instanton}
To count fermion zero modes one recasts the
equations of motion for fermions living on the M5-instanton as
a set of equations on differential forms on the divisor $D$.
For this purpose we  recall below  how
world-volume fermions transform under the rotations of the normal and
tangent directions.
The normal bundle to the M5-brane has a product form $\mathbb{R}^3 \times {\cal N},$ where $\mathbb{R}^3$ stands for external space\footnote{ For computation of instanton generated superpotential
we work in Euclidean signature in external 3D space. }
and ${\cal N}$ is the line bundle describing one complex normal 
direction inside $Y_4.$ Let us introduce complex coordinates $z^i,\quad i=1,2,3$ along the
divisor $D$ and the complex coordinate $w$ normal to $D$ inside $Y_4.$ 

The fermions 
$\theta=\begin{pmatrix}\theta_{\a}^{ A +} \cr \theta_{\a}^{ A  -} 
\end{pmatrix}$
living on the M5-brane transform in the representation 
${\bf 4 \otimes 2 \otimes 2}$ 
under $Spin(6)\times SO(3)\times SO(2)$. 
Here $A=1,2$ is a spinor under  
external $SO(3),$ the $+(-)$ stands for a chiral(anti-chiral)
spinor of SO(2)  and $\a=1,\ldots,4$  is a chiral
spinor  of Spin(6).
Now we use the  fact (see for example  \cite{Witten:1996bn}) that
the bundle $S^+$ of chiral spinors on a K\"ahler manifold
of complex dimension three is isomorphic to the bundle  
\be
\Bigl( \Omega^{(0,0)}\otimes K^{\half}\Bigr) \oplus
\Bigl(\Omega^{(0,2)}\otimes K^{\half}\Bigr)\,.
\ee
Here $\Omega^{(0,p)}$ stands for the bundle of $(0,p)$ forms.
 We will further use that 
the normal bundle on the divisor in $Y_4$
is isomorphic to the canonical bundle $K.$ 
Recalling that  $\theta$ is a section  
 of the bundle\footnote{Here we are not writing explicitly spinor index  in $\mathbb{R}^3$.} $S^+\otimes K^{\half} \oplus S^+\otimes K^{-\half},$
we find the following degrees of freedom.   
A (0,2) form $a_{(2)}^w$ taking values in the canonical bundle $K$,
a section of $K$,  $a_{(0)}^w $, as well as
a (0,2) form $b_{(2)}$
and a  scalar $b_{(0)}. $

Locally we write $\theta$ in terms of these degrees of freedom
as follows 
 \be \label{forms} \theta=\bigl( a_{(0)}^w +a_{{\bar i} {\bar j}}^w\gamma^{\bar i} {\tilde \gamma}^{\bar j}\bigr)T_{w}\e+ 
 \bigl( b_{(0)} +b_{{\bar i} {\bar j}}\gamma^{\bar i} 
{\tilde \gamma}^{\bar j}\bigr)\e \,,
\ee
where the  chiral spinor $\e$ satisfies 
\be \label{chir_spin}{\tilde \gamma}^i\e=0,\quad i=1,2,3,\quad T_{\bar w} \e=0 \,.
\ee
Here $T_w,T_{\bar w} $ are $SO(2)$ Dirac matrices:
\be \label{sotwo}T_{w} T_{\bar w} + T_{\bar w} T_{w}=2g_{w \bar w}.\ee
Meanwhile, the six dimensional chiral(anti-chiral) gamma matrices ${ \tilde \gamma}_i,
\, { \tilde \gamma}_{\bar j}(\gamma_{i},\, \gamma_{{\bar j}})$ have the properties
\be \label{sixdim}
\gamma_{\bar j}{\tilde \gamma}_i+ \gamma_{i}{\tilde \gamma}_{\bar j}=2g_{i \bar j}  \,,
\ee
where $g_{i \bar j}$ is K\"ahler metric on the divisor $D$.

The equation of motion for fermions reads \cite{Kallosh:2005yu}
\be \label{ferm}
{\tilde \gamma}^i \nabla_i \theta + {\tilde \gamma}^{\bar i} 
\nabla_{\bar i} \theta -
{ 1\over 8}T_{\bar w} {\tilde \gamma}^{{\bar i} {\bar j} k}
G_{{\bar i} {\bar j} k}^{~~~\bar w}\theta
-{ 1\over 8}T_{w} {\tilde \gamma}^{i j {\bar k} }
G_{ i j \bar{k}}^{~~~{w}}\theta=0 \,.
\ee
Here the covariant derivatives $\nabla_{j},\nabla_{\bar j}$ include 
the connection on the bundle of chiral Spin(6) spinors
 as well as connection on the spin bundle
derived from  the normal bundle ${\cal N}$.

Using (\ref{forms}) we find from (\ref{ferm})
 the following set of equations\footnote{
$X_{[\bar i_1 \ldots \bar i_p]}={1 \over p!}\bigl( X_{\bar i_1 \ldots \bar i_p} \pm \hbox{permutations}\bigr)$}:
\be\label{geom} 
\ba
\p_{[\bar i} b_{{\bar j}{\bar k}]}=&0\cr
 4 \p^{\bar j}b_{\bar j \bar k}+\p_{\bar k} b_{(0)}=&0  \cr
D_{[\bar m_1} a_{\bar m_2 \bar m_3]}^w=&0 \cr
4D^{\bar j} a_{\bar j \bar k}^w+ D_{\bar k}a_{(0)}^w=& -G^{{\bar i} {\bar j}~~w}_{~~~\bar k } b_{\bar i \bar j} \,.
\ea
\ee
In the equations (\ref{geom}) the covariant
differentials include  connection on the canonical
bundle and  we used the primitivity condition on $G$-flux.

There are always universal modes counted by
$2\times h^{(0,0)}(D)=2$,
where the factor of $2$ accounts for two spinor indices in $R^3.$
 Non-universal modes without $G$-flux are counted by
\begin{equation}2\times \left[h^{(0,1)}(D)+h^{(0,2)}(D)+h^{(0,3)}(D)\right].\end{equation}

It is further known \cite{Saulina:2005ve, Kallosh:2005gs, Lust:2005cu, Tsimpis:2007sx}
that modes counted by $h^{(0,2)}(D)$ may
be lifted by G-flux since the G-flux couples to them directly in \eqref{geom}.  
An especially simple  case
of such lifting occurs \cite{Saulina:2005ve} if $h^{(0,1)}(D)= h^{(0,3)}(D)=0$
and $h^{(0,2)}(D)=1.$ 

\subsubsection{$h^{(0,k)}(D)$ for $D=p^*S_2$}
To study the zero mode structure of an M5 then we must be able to compute the Hodge numbers $h^{(0,k)}(D)$ for a vertical divisor $D=p^*S_2$.  From the connection to D3-instantons, one intuitively expects that these should be related to cohomologies of the normal bundle of $S_2$ inside $B_3$.  We describe this relation for the special case where $S_2$ is Hirzebruch $\mathbb{F}_n$ or del Pezzo $dP_m$ in the following.

Let $P: D \rightarrow S_2$ be the elliptic fibration arising
as a restriction of $\pi: Y_4\rightarrow B_3.$ 
We now use the Leray spectral sequence to relate the cohomology of the divisor $D=\pi^*S_2$ to the cohomology of $S_2$.  We start from $E_2^{ij}$ with
$i=0,1,2$ and $j=0,1$, where each entry $E_2^{ij}$ is given by $H^i(S_2, R^j P_*\mathcal{O}_D)$
\be
E_2^{ij}: \qquad 
\begin{array}{ccc}
H^0(S_2,R^1P_* \mathcal{O}_D)&H^1(S_2,R^1P_* \mathcal{O}_D) & H^2(S_2,R^1P_* \mathcal{O}_D) \cr
&&\cr
H^0(S_2,\mathcal{O}_{S_2})&H^1(S_2,\mathcal{O}_{S_2}) & H^2(S_2, \mathcal{O}_{S_2}) 
\end{array}
\ee
We will always take $S_2$ to be Fano $\mathbb{F}_n$ or del Pezzo $dP_m$ so
\be
H^1(S_2,\mathcal{O}_{S_2}) =H^2(S_2, \mathcal{O}_{S_2}) =0\,.
\ee
Then $E_{\infty}^{i,j}=E_2^{i,j}$ and
\be
h^{(0,0)}(D)=1,\quad h^{(0,k)}(D)=h^{k-1}(S_2,R^1P_* \mathcal{O}_D) \,,\quad k=1,2,3.
\ee
Now the key step is to relate $R^1P_* \mathcal{O}_D$ with the normal bundle of $S_2$
in $B_3$: 
\be
R^1P_* \mathcal{O}_D=N_{S_2\vert D}=K_{B_3}\vert_{S_2}=
K_{S_2}\otimes N^{-1}_{S_2\vert B_3}.
\ee
So that using Serre duality
\be\ba\label{h0p}
h^{(0,1)}(D)&=h^2(S_2,N_{S_2\vert B_3}) \cr
h^{(0,2)}(D)&=h^1(S_2,N_{S_2\vert B_3}) \cr
h^{(0,3)}(D)&=h^0(S_2,N_{S_2\vert B_3}) \,.
\ea\ee
The absence of non-universal fermi zero modes therefore corresponds to the vanishing of suitable cohomologies of the normal bundle of $S_2$ inside $B_3$.

\subsection{$G$-flux, the M5 path-integral, and charged couplings}
\label{subsec:M5Gflux}

Let us now turn to the effect of $G$-flux on the M5-instanton partition function and the corresponding generation of charged couplings.  For this, we only really need to consider the topological couplings:
\begin{equation}S_{M5} \sim i\tau_{M5}\int_{M5}\left(C_6-\half b_2\wedge G_4+\ldots\right) \,.
\label{M5act}\end{equation}
Here, $b_2$ is the chiral 2-form while $C_6$ and $G_4$ are restrictions of the corresponding 11d fields to the M5 worldvolume.  The form of the second coupling is essentially determined by the fact that $b_2$ and $C_6$ shift under the bulk gauge transformation $C_3\rightarrow C_3+d\Lambda_2$ according to
\begin{equation}
b_2\rightarrow b_2 + \Lambda_2\,,\qquad C_6\rightarrow C_6 +\half \Lambda_2 \wedge G_4 \,.
\end{equation}
Once the second coupling in \eqref{M5act} is included the sum is nicely gauge invariant.  One useful feature of \eqref{M5act} is that it makes clear how the physics of our M5-instanton is impacted by a nontrivial $G$-flux: the $G$-flux introduces a source for the 2-form $b_2${\footnote{{We are grateful to S.~Sethi for emphasizing to us the important role played by the chiral 2-form and for helpful discussions about this point.}}}.  This, in turn, means that classical solutions for $b_2$ correspond to solitons that we should interpret as M2 branes ending on the M5 \cite{Howe:1997ue}.  In the case at hand, our M5 is Euclidean so the M2s will have two spatial directions parallel and a timelike direction transverse to the M5.

In principle, if we know the $G$-flux then we can determine exactly what M2-brane configuration will be described by the solitonic solutions to the equations of motion for $b_2$.
It is easier to think not in terms of solitons of the M5 worldvolume theory, however, but to instead introduce M2s directly into the game in order to cancel the flux-induced tadpole.  Each M2 has a worldvolume coupling to $C_3$ along with a boundary coupling to the 2-form $b_2$ \cite{Bergshoeff:1987cm, Bergshoeff:1987qx}
\begin{equation}S_{M2}\sim i\tau_2\left(\int_{M2}C_3 - \int_{M2\cap M5} b_2+\ldots\right)\,.
\label{M2act}\end{equation}
The idea now is to include M2-branes in such a way that the coupling to $b_2$ in \eqref{M2act} exactly cancels the one induced by the $G$-flux in \eqref{M5act}.  The resulting action for $b_2$ will be the ordinary one that we would have if no $G$-flux were present at all.  Note that the individual gauge invariance of \eqref{M5act} and \eqref{M2act} guarantees that the sum is gauge invariant as well.  When the source terms for $b_2$ in \eqref{M5act} cancel off the boundary couplings in \eqref{M2act}, then this guarantees that the net charge of the individual M2-branes under a bulk gauge transformation, as determined by the shifts in $i\tau_2\int_{M2}C_3$, is precisely canceled by the charge of the M5-instanton, as determined by the shift in $i\tau_5\int_{M5}C_6$.  Invariance under bulk gauge transformations is thus automatic and hence invariance under any $U(1)$s of the type considered in section \ref{subsec:dp9U1} is automatic as well.

To see which M2-branes we have to add, let us rewrite the $G_4$-dependent part of \eqref{M5act} as
\begin{equation}-\frac{i}{2}\tau_5\int_{M5}b_2\wedge G_4 = -\frac{i}{2}\tau_5\int_{PD_{M5}[G_4]}b_2\,,\end{equation}
where $PD_{M5}[G_4]$ is the Poincare dual of $G_4$ inside the M5.  A collection of M2-branes that cancels this is one for which the curves $\Sigma_i=M5\cap M2_i$ satisfy
\begin{equation}\sum_i \tau_2\Sigma_i = -\half \tau_5 PD_{M5}[G_4] \,.
\label{PDG}\end{equation}
The M2s that participate in our instanton-generated coupling are determined by the Poincare dual of the $G$-flux inside the M5 worldvolume.  Note that the gymnastics we have just described is nothing more than a procedure for describing how to associate a collection of Wilson surface observables of the type considered in \cite{Donagi:2010pd} with a charged coupling involving 4-dimensional fields associated to a particular collection of wrapped M2's.  Our description in terms of \eqref{PDG} is particularly convenient because our construction of $G$-flux is one in which we essentially define it by the Poincare dual.  Before going too far, though, we have to be somewhat careful about normalization.  When we build $G$-flux as in section \ref{subsec:specdiv} by specifying an integer combination of holomorphic surfaces ${\cal{G}}$ inside the ``Tate divisor", that flux is quantized according to
\begin{equation}G_{\rm Tate}+\frac{1}{2}c_2(Y_4)\in H^4(Y_4,\mathbb{Z}) \,.
\label{ourquantization}\end{equation}
In terms of the 3-form $C_3$ appearing in the action above, this $G_{\rm Tate}$ is then really{\footnote{This can be determined in the usual way by looking at the phase of an M2 as it moves around in a flux background \cite{Witten:1996md}.}}
\begin{equation}G_{\rm Tate} = \frac{2\pi}{\tau_2}dC_3\,,\end{equation}
so that, using the M2 and M5 tensions $\tau_2=(2\pi)^{-2}$ and $\tau_5=(2\pi)^{-5}$, we have that
\begin{equation}\sum_i \Sigma_i = -\frac{1}{2}PD_{M5}[G_{\rm Tate}] \,.
\label{relate}\end{equation}
Up to a factor $(-\frac{1}{2})$, then, the surface that we use to describe the $G$-flux tells us precisely which wrapped M2-brane states participate in the instanton-generated coupling.

We still have to discuss the nature of $PD_{M5}[G_{\rm Tate}]$ to get our prescription for determining the charged coupling generated by an M5 instanton.  Before getting to that, though, we should say something about the puzzling factor of $-\frac{1}{2}$.  This seems to be the statement that $G$-flux can induce a half-integral M2 charge on the M5 worldvolume.  We are not aware of any reason for this charge failing to be integer quantized and, correspondingly, do not know of any ways to suitably cancel it.  It seems, then, that a vertical divisor $\pi^*S_2$ to which the restriction of $G_{\rm Tate}$ is an odd class simply cannot house an M5 instanton.  We would be very interested in a better understanding of this issue but unfortunately do not have anything substantive to say about it at the moment.

\subsubsection{The nature of $PD_{M5}[G_{\rm Tate}]$}

Let us now consider how the 2-cycles specified by $PD_{M5}[G_{\rm Tate}]$ can be determined from the geometric construction of $G$-fluxes in section \ref{subsec:specdiv}.  The $G$-flux there was actually constructed as a difference
\begin{equation}G_{\rm Tate} = {\cal{G}}-G_0 \,,
\end{equation}
where ${\cal{G}}$ is a surface inside the ``Tate divisor", ${\cal{C}}_{\rm Tate}$, and $G_0$ is a class in $Y_4$ that we must subtract to ensure that $G_{\rm Tate}$ is orthogonal to horizontal and vertical divisors.  The restriction of ${\cal{C}}_{\rm Tate}$ to an M5 worldvolume will be a spectral surface ${\cal{C}}_{\rm Tate}|_{M5}$ in that worldvolume while ${\cal{G}}$ will be a curve $\Sigma_{\cal{G}}$ that sits inside it.  The restriction $G_0|_{M5}$ is also a curve $\Sigma_{G_0}$ inside the M5 but will not lie in the spectral surface ${\cal{C}}_{\rm Tate}|_{M5}$.  The object $PD_{M5}[G_{\rm Tate}]$ is then given by the difference
\begin{equation}PD_{M5}[G_{\rm Tate}]=\Sigma_{\cal{G}}-\Sigma_{G_0}\end{equation}
in the resolved geometry.

To get a better feel for $PD_{M5}[G_{\rm Tate}]$, let us look at its intersections with divisors inside the M5.  Any intersection of $\Sigma_{\cal{G}}$ with a horizontal or vertical divisor will  be canceled by a corresponding contribution from $\Sigma_{G_0}$.  The M5 may however contain additional divisors upon resolution, if it contains a curve of singularities $\Sigma_{sing,i}$.  Associated with that curve of singularities is a matter surface `$\Sigma_{sing,i}\times C_i$' that will represent a divisor in the M5 where the intersection with $\Sigma_{\cal{G}}$ is not canceled by a contribution from $\Sigma_{G_0}${\footnote{Note that $\Sigma_{\cal{G}}$ will generically intersect $\Sigma_{sing,i}$ inside the M5 even though they are both codimension 2.  This is because both curves actually sit inside the spectral surface ${\cal{C}}_{\rm Tate}|_{M5}$, with respect to which each is codimension 1.  Let us also emphasize that it is not enough for the M5 to contain isolated points of singularities.  Even though $\Sigma_{\cal{G}}$ and an isolated point singularity both sit inside ${\cal{C}}_{\rm Tate}$, they will generically not intersect.  Said differently, an isolated point singularity inside the M5 does not give rise to a new divisor in the M5 upon resolution over which we can integrate $G_{\rm Tate}$.}}.  Divisors of this type represent the only ones in M5 over which $G_{\rm Tate}$ can give a nonzero integral.  If the M5 does not contain any curves of singularities, then, $G_{\rm Tate}$ will have a trivial restriction to its worldvolume.
Given our geometric description of $G_{\rm Tate}$, the curve $PD_{M5}[G_{\rm Tate}]$ is actually easy to describe.  We have
\begin{equation}PD_{M5}[G_{\rm Tate}] = \sum_i n_i C_i \,,
\end{equation}
where $C_i$ is a degenerate cycle and $n_i$ is the intersection number $\Sigma_{\cal{G}}\cdot_{ {\cal{C}}_{\rm Tate}|_{M5}} \Sigma_{sing,i}$ with $\Sigma_{sing,i}$ the curve of singularities inside ${\cal{C}}_{\rm Tate}|_{M5}$ along which $C_i$ degenerates.

We can say all of this in a much simpler way.  We constructed $G_{\rm Tate}$ so that it would correspond to a (2,2)-form in the resolved fourfold that has nonzero integral only over matter surfaces, that is to say surfaces comprised of curves inside $Y_4$ and 2-cycles that degenerate along those curves.  The restriction of $G_{\rm Tate}$ to an M5 instanton will therefore only be nontrivial if the instanton contains at least one of these matter surfaces in its entirety since otherwise there will not be surfaces over which we can integrate $G_{\rm Tate}$ to get a nonzero answer.  Now, the M5 wraps a divisor in $Y_4$ while the singularities of interest lie along curves in $Y_4${\footnote{It is true that the M5 intersects surfaces of singularities (like the one above $S_{\rm GUT}$) in curves.  $G$-fluxes that can have nonzero integral over these singularities are precisely fluxes that break the corresponding non-Abelian gauge group, i.e. worldvolume flux associated to the stack of 7-branes.  Hypercharge flux is an example of such a flux so this will be an important issue for model-building applications (though trivializable hypercharge fluxes should not pose a problem because the curves in question are homologically nontrivial in $B_3$ as the intersection of a surface $S_{\rm GUT}\subset B_3$ with the divisor $S_2\subset B_3$ that is the base of the M5-instanton).  For simplicity, we focus only on bulk $G$-fluxes in this paper.}}.  The generic situation, then, is for the M5 to intersect a curve of singularities only in isolated points.  This means that bulk $G$-fluxes of the type described in section \ref{subsec:specdiv} generically have trivial restriction to M5 instantons.  Based on our previous discussions, this means that the topological term is absent from the M5-instanton action and there is no source term for $b_2$.  This removes one obstacle to obtaining a nonzero coupling, leaving only the structure of fermi zero modes to be checked.  The classical action of such an M5-instanton is invariant under all bulk gauge transformations so this is consistent with all of the usual $U(1)$ selection rules.

\section{Instanton induced superpotential: an Example}
\label{sec:application}

In this paper so far we
have developed the general framework for studies of global $G$-fluxes and $U(1)$ symmetries, and gave criteria
when instanton contributions to the superpotential arise. We will next apply these in a concrete example, with the base threefold 
that we constructed in \cite{Marsano:2009ym}. 
The construction of $B_3$ in our earlier work \cite{Marsano:2009ym} relied on a chain of blowups along a nodal curve in $\mathbb{P}^3$, and the final threefold was 
obtained after a flop transition. There is a slightly more elegant way to reach this geometry, which does not require a flop, that effectively reverses the order of the blow-ups.  We only give a heuristic description of this approach in the main text below but a detailed description can be found in appendix \ref{app:ThreeFold}.


\subsection{Divisors that contribute superpotential couplings}
\label{subsec:Divs}

The first obstruction to obtaining superpotential couplings from an M5-instanton is the fermi zero mode structure.  As we discussed in section \ref{subsec:fermimodes}, obtaining the right structure for a vertical divisor of the form $\pi^*S_2$ requires the cohomology groups $H^p(S_2,N_{S_2/B_3})$ to be trivial for $p=0,1,2$.  As a start, then, we should look for divisors $S_2$ in $B_3$ that are not moveable so that at least the $p=0$ condition is satisfied. More precisely, the relevant divisors $D= \pi^*S_2$ have to satisfy
$h^{(0,3)}(D) = h^0(S_2, N_{S_2|B_3}) =0$.
Before listing candidate divisors in the threefold of \cite{Marsano:2009ym} that satisfy this criterion, though, let us quickly review its basic structure.

We can obtain the threefold of \cite{Marsano:2009ym} by starting with $\mathbb{P}^3$ and picking out a distinguished plane cubic curve $C_{nodal}$ with a node.  Because of the node, we can think of $C_{nodal}$ as a pinched torus obtained by gluing together the north and south poles of a $\mathbb{P}^1$.  The construction of $B_3$ now 
proceeds in two steps:

\begin{itemize}
\item First, blow up $\mathbb{P}^3$ at the node
\end{itemize}

This step separates the two branches of $C_{nodal}$ so that its proper transform, $C_{nodal}'$, is a smooth $\mathbb{P}^1$.  The exceptional divisor of this blow-up, $E$, is a $\mathbb{P}^2$ that meets $C_{nodal}'$ in exactly two points.  Now, onto the second step

\begin{itemize}
\item Second, blow up along $C_{nodal}'$
\end{itemize}

Because $C_{nodal}'$ is a $\mathbb{P}^1$, the exceptional divisor $D'$ of this second blow-up is a Hirzebruch surface $\mathbb{F}_n$ where $n$ is determined by the normal bundle of $C_{nodal}'$.  It is easy to show, as we do in Appendix \ref{app:ThreeFold}, that $D'$ is in fact an $\mathbb{F}_4$.  Note that since $E$ met $C_{nodal}'$ in exactly two points, the preimage $E'$ of $E$ under the second blow-up is a $dP_2$ surface.  Further, $D'$ meets $E'$ in a sum of two curves.  From the perspective of $E'$, these are the two exceptional curves of the $dP_2$.  From the perspective of $D'$, these curves are both in the fiber class of the Hirzebruch.  As a result, the two exceptional curves of $E'$ are in fact homologous to one another in $B_3$ by a 3-chain that sits inside $D'$.  This construction can easily be generalized to engineer threefolds where $E'$ is replaced by a $dP_n$ surfaces provided we replace $C_{nodal}$ with a curve that has a higher degree singularity in place of the node.

\begin{figure}
\begin{center}
\epsfig{file=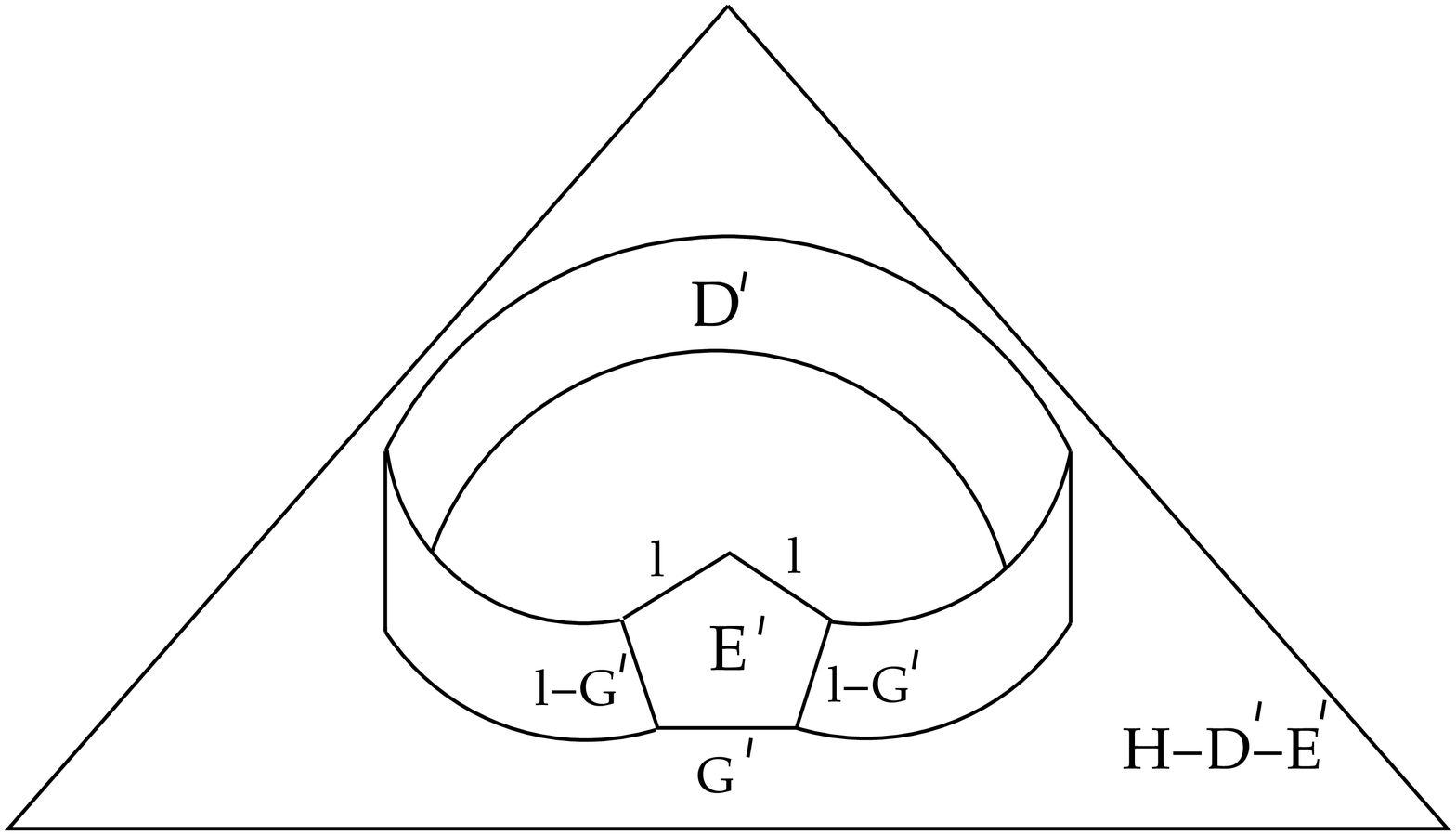,width=0.6\textwidth}
\caption{Cartoon of the threefold $B_3$ from \cite{Marsano:2009ym}.  When $D'$ and $E'$ are blown down we are left with a nodal cubic that lies in the plane $H-D'-E'$.  $\ell$ and $G'$ label curve classes in the notation of \cite{Marsano:2009ym}.}
\end{center}
\label{fig:cartoon}
\end{figure}

We reproduce the cartoon of $B_3$ from \cite{Marsano:2009ym} in figure \ref{fig:cartoon}.  From its construction, we see that there are two obvious divisors for which $h^0(S_2,N_{S_2/B_3})=0$
\begin{equation}\begin{split}E' &\cong dP_2 \\
D' &\cong \mathbb{F}_4 \,.
\end{split}\label{twodivisors}\end{equation}
All other divisors descend from polynomials in our original $\mathbb{P}^3$ that have specified intersections with the original nodal curve, $C_{nodal}$, or its node.  Almost all of these are obviously deformable but there is one more candidate that is not: the specific hyperplane of $\mathbb{P}^3$ in which our plane curve $C_{nodal}$ sat.  The proper transform of this hyperplane is the divisor
\begin{equation}H-D'-E'\cong dP_1 \,.
\label{thirddivisor}\end{equation}
Indeed, the hyperplane becomes a $dP_1$ after the first blow-up that is unaffected by the second blow-up.

So, in total, we have three candidate divisors in $B_3$ that can serve as the base of M5 instantons with the right zero mode structure.  In our discussion above, though, we have only argued that $H^0(S_2,N_{S_2/B_3})$ is trivial for each candidate.  It remains to check $H^p(S_2,N_{S_2/B_3})$ for $p=1,2$.  We do this in Appendix \ref{app:ThreeFold} with the following results.  All of the required cohomologies vanish for $E'$, $H-D'-E'$, and $D'$ so that the fermi zero mode structure is exactly right in each of these cases.  

Let us turn now to the restriction of $G$-flux.  In general, $D'$ and $H-D'-E'$ will not contain any entire curves of singularities.  As discussed in section \ref{subsec:M5Gflux}, this means that the $G$-flux used to engineer a chiral spectrum restricts trivially.  Provided no other $G$-fluxes have nontrivial restriction, both $\pi^*D'$ and $\pi^*(H-D'-E')$ can be expected to yield superpotential couplings that can be used for K\"ahler moduli stabilization.

What about $E'$?  This divisor is special because it is exactly the GUT divisor on which the 7-branes are wrapped in $SU(5)_{\rm GUT}$ models based on the geometry of \cite{Marsano:2009ym}.  Any M5 wrapping $\pi^*E'$ will therefore contain several curves of singularities.  The restricted $G$-flux will be nontrivial in general so the M5 wrapping $\pi^*E'$ (which is really just capturing the contribution of a gauge theory instanton of the 7-brane worldvolume theory) will not play any role in K\"ahler moduli stabilization{\footnote{In three generation models, in fact, the Poincare dual of the $G$-flux inside $\pi^*E'$ should be odd.  In this case, we are led to expect that a single M5 wrapping $\pi^*E'$ does not contribute any superpotential coupling at all.}}.

\subsubsection{Any hope for K\"ahler moduli stabilization?}

Since only two divisors in $B_3$ can lead to superpotential corrections from M5-instantons, it may be difficult to stabilize K\"ahler moduli.  This is because we have 3 K\"ahler moduli, which can be parametrized by writing the divisor corresponding to the K\"ahler form as
\begin{equation}J = mH - aE' - bD'\,.
\end{equation}
Explicitly, the K\"ahler cone is easy to determine and corresponds to $m$, $a$, and $b$ satisfying
\begin{equation}a>2b>0,\qquad m>a+b \,.
\end{equation}
While it is not always true that we need as many nonperturbative terms in the superpotential as we have K\"ahler moduli \cite{Bobkov:2010rf}, we believe that instanton corrections from $D'$ and $H-D'-E'$ in the threefold $B_3$ are not enough by themselves to stabilize everything.  Fortunately, it might be possible to engineer a little help from an unexpected place.  {The geometry of \cite{Marsano:2009ym} admits a nice divisor class that does not intersect the GUT divisor $S_{\rm GUT}=E'$.  This class, which we call $H$ because it descends from a hyperplane in $\mathbb{P}^3$ that does not contain the node of $C_{nodal}$, is somewhat like the divisor that supports the other $E_8$ in models with Heterotic dual.  If we can manage to engineer a hidden sector gauge group on $H$ that undergoes gaugino condensation, the resulting superpotential correction can be used as the third nonperturbative term that we need to stabilize all of the K\"ahler moduli.}


\subsubsection{Hidden Sector}

{We therefore propose that models based on the geometry of \cite{Marsano:2009ym} should, in the future, realize a hidden sector on $H$ and study the interplay of global fluxes with that sector to determine the presence or absence of gaugino condensation.  Note that even the realization of a hidden sector is nontrivial; it is relatively easy to see that a generic $SU(5)_{\rm GUT}$ model based on \cite{Marsano:2009ym} will not have surfaces of singularities anywhere away from $S_{\rm GUT}$.  When trying to engineer a surface of singularities along the divisor $H$, one must take some care to make sure that non-minimal singularities are not introduced.  We now proceed to describe one example that implements a hidden sector.  In the future, we hope to report on model-building efforts that engineer hidden sectors in more realistic models while implementing the approach to $U(1)$s and global fluxes described earlier in this paper.}

{To get a simple example of a hidden sector on $H$, let us start with the global Weierstrass models studied in \cite{Marsano:2009ym}.  There we engineered an $SU(5)_{\rm GUT}$ singularity at $z=0$ (on $S_{\rm GUT}=E'$) by specifying a rather general Tate form.  Expressed in the notation of \cite{Andreas:2009uf,Marsano:2009ym} this is}
\be
y^2 = x^3 + z^5 g_5+\frac{1}{48} f_3 x z^3+h x y -\frac{1}{4} H x^2 z+\frac{1}{12} q  z^2 y \,.
\label{2009tate}\ee
{Here,} the global sections $g_5, f_3, h, H, q$ are rescalings of $a_m$ in the global Tate model. {We use} $P=  - 3 H q^2 - 3 h f_3 q + 2 g_5 h^2$ {to denote} the section that gives rise to the codimension two enhancement to $SU(6)$. 
The holomorphic sections on $B_3$ {that we use to build the sections in \eqref{2009tate} are}
\be\label{SectionsX}
\begin{array}{|l||l| }\hline
\text{Holomorphic Section} & \hbox{Divisor class} \cr\hline\hline
Z_4 & H\cr\hline
Z_{1,2} & (H-E') + E' =H \cr\hline
Z_3 & (H-D'-E') + (D'+E')= H \cr\hline
W_{1,2,3} & H-E'\cr\hline
W_4 & 3H -D'-2E' \cr\hline
V_1 & ( 3H - D' -2E') + E' = 3 H - D' - E' \cr\hline
V_0 & H-D'-E'\cr\hline
{z} & {E'} \cr\hline
\end{array}
\ee
{We actually never use $Z_{1,2,3}$ or $V_1$ explicitly.  The divisors defined by each of these are reducible which indicates that each is the product of more elementary sections.}
Consider the ansatz for the section $h$ {of \eqref{2009tate}} which is in the class $4 H - D' - 2 E'$
\be
h = (Z_4 + z P_1 ( W_1, W_2, W_3)) (W_4 - (Z_4 + z P_1(W_1, W_2, W_3) )V_0 A_1(W_1, W_2, W_3)) \,,
\ee
where $A_1$ and $P_1$ are  linear polynomials in $W_i$, $i= 1, 2, 3$\footnote{Note that in the examples in appendix E of \cite{Marsano:2009ym} we considered a slightly less general ansatz, where $P_1=0$. }. Making the same choices otherwise for $P,H,  q$ and  $f_3 =0$ and  $g_5(P, H, q)$ determined in terms of the sections $P, H, q$ as in example 1 of appendix E in \cite{Marsano:2009ym}, we arrive at an elliptic fibration, which automatically realizes the $SU(5)_{\rm GUT}$ singularity at $z=0$ with matter {curves} and Yukawas. 

To determine whether there is any nonabelian gauge enhancement along the divisor $H$, we expand the discriminant around $Z_4 =0$. For $P_1 =0$ the discriminant has vanishing order $\Delta \sim Z_4^{12}$ and in fact, can be brought into non-minimal Tate form, which is disfavored. We therefore consider $P_1$ a generic polynomial in $W_{1, 2, 3}$, which yields
\be
\Delta  = Z_4^6 \tilde{\Delta} + O(Z_4^7) \,,
\ee 
where $\tilde{\Delta}$ has vanishing order in the other sections of the order $z^{12}$, $V_0^6$, $W_i^6$ and $W_4^0$ {\footnote{{We do not  worry about the high degree of $z$ in $\tilde{\Delta}$ since $z=Z_4=0$ admits no solutions.}}. After shifting $x, y$ by $Z_4$, the Weierstrass form can be brought globally into Tate form for $I_6$  without monodromy, which  is $SU(6)$ \cite{Bershadsky:1996nh, Katz:2011qp}, i.e.
\be
y^2 +a_1xy+a_3y=x^3 +a_2x^2 +a_4x+a_6 \,,
\ee
where
\be
a_1 = O(Z_4^0) \,,\quad
a_2 = O(Z_4^1) \,,\quad
a_3 = O(Z_4^3) \,,\quad
a_4= O(Z_4^3) \,,\quad
a_6 = O(Z_4^6)\,.
\ee
So in summary this exemplifies that a non-abelian gauge group can reside on the divisor $H$.

\subsection{Superpotential  and SUSY vacua}

Thus far, we have argued that geometries based on the threefold of \cite{Marsano:2009ym} will exhibit superpotential couplings from M5-instantons wrapping $\pi^*D'$ and $\pi^*(H-D'-E')$.  Provided we get help from a gaugino condensate on $H$, this should be enough to stabilize K\"ahler moduli.  In this subsection, we investigate the potential that would result and show that it indeed exhibits a supersymmetric minimum.


Concretely, the superpotential of interest is 
\begin{equation}
W_{non-pert} = \sum_{D\in \{D',H-D'-E',H\}} a_D e^{-T_D} \,.
\label{Wnonpert}\end{equation}

To proceed, we should write the real parts of the $T_D$'s of interest in terms of the K\"ahler moduli.  Recall that we parametrize the K\"ahler form as
\begin{equation}J=mH-aE'-bD'\end{equation}
with the K\"ahler cone specified by
\begin{equation}a>2b>0,\qquad m>a+b\,.
\end{equation}


The divisors of interest that give contributions from wrapped instantons are \eqref{twodivisors} and \eqref{thirddivisor} and
have volumes given by 
\begin{equation}\begin{aligned}\text{Re }T_{D'}&=  2b(3m-2a-3b) \cr
\text{Re }T_{H-D'-E'} &= (m-a-b)(m+a-5b) \cr
\text{Re} T_{H}  &= m^2 - 3 b^2 \,.
\end{aligned}\end{equation}
Furthermore, the K\"ahler potential is 
\begin{equation}K = -2\ln J^3 = - 2\ln\left(m^3-a^3+6b^3-9mb^2+6ab^2\right) \,.
\end{equation}
The superpotential contribution from M5 instantons is then
\be
W= a_1  e^{-T_{H- D'-E'}}+a_2e^{-T_{D'}}+ a_3e^{-T_H}  + W_0 \,.
\ee
We will in the following set $a_1=1$ without loss of generality.
The scalar potential is 
\begin{equation}V = e^{K}\left(K^{a\bar{b}}(D_aW)(D_{\bar{b}}W) - 3|W|^2\right) \,,
\end{equation}
where $D_a$ is the usual K\"ahler covariant derivative
\begin{equation}D_a f = \partial_a f + (\partial_aK)f \,.
\end{equation}
To find SUSY vacua inside the K\"ahler cone, consider the three equations
\begin{equation}
D_{a, b, m} W =0  \,.
\label{susyeqns}\end{equation}
In these equations, we follow convention and absorb the axion vevs into the coefficients $a_D$.  These vevs are ultimately determined by the phases of the $a_D$.
Instead of looking for solutions for $m$, $a$, and $b$ in terms of the superpotential parameters, we instead solve \eqref{susyeqns} for $W_0$, $a_2$, and $a_3$ as functions of the K\"ahler moduli.  That way, given a particular place in the K\"ahler cone, we can determine the precise superpotential couplings that give rise to a supersymmetric vacuum there.  Doing this, we find
\be
\ba
W_0  =& -\frac{e^{(a+b-m) (a-5 b+m)} (3 (a-b+m)+\text{Vol})}{3 a}
\ea\ee
where 
\be
\text{Vol} = -a^3+6 a b^2+6 b^3-9 b^2 m+m^3 \,,
\ee
as well as 
\be
\ba
a_2=&\frac{(a-b) e^{a^2-8 a b-11 b^2+12 b m-m^2}}{a} \cr
a_3=&\frac{(m-a) e^{a^2-4 a b-8 b^2+6 b m}}{a} \,.
\ea
\ee
From these expressions it is clear that we can achieve supersymmeric solutions inside the K\"ahler cone with $a_2, a_3$ (and $a_1$) all positive and $W_0<0$.


\section*{Acknowledgements}

We would like to thank I.~Garc\'{i}a-Etxebarria, J.~Halverson, W.~Taylor, and especially S.~Sethi for valuable discussions.  We also thank M.~Wijnholt for helpful discussions on the first draft of this paper and M.~Esole for pointing out an error in the computation of bundle cohomologies on $\mathbb{F}_4$ in Appendix \ref{app:subsubsecF4} in v1.  We are all very grateful to the Caltech theory group for their generous hospitality at a crucial stage of this work.  JM would also like to thank the Perimeter Institute for Theoretical Physics, the theoretical physics and algebraic geometry groups at The Ohio State University, the theory group at the University of Pennsylvania, the organizers of the Spring 2011 String Vacuum Project meeting, and the Center for Theoretical Physics at the Massachusetts Institute of Technology for their hospitality at various stages of this work.  JM and NS are grateful to the organizers of the Fall 2010 String Vacuum Project at The Ohio State University during the initial stages of this work.  NS is grateful to the University of Chicago theory group for hospitality. SSN thanks the organizers of StringMath 2011 for hospitality at UPenn and a stimulating conference. 
 The work of JM is supported by DOE grant DE-FG02-90ER-40560 and NSF grant PHY-0855039. 
 We thank Kolya Gromov for communication assistance during the final stages of this work.


\newpage
\startappendix


\section{Three-fold}
\label{app:ThreeFold}

In this appendix we explain an alternative construction of the threefold of  \cite{Marsano:2009ym}, which was already outlined in section \ref{subsec:Divs}.  We start with $\mathbb{P}^3$ parametrized by homogeneous coordinates $[Z_1,Z_2,Z_3,Z_4]$ and blow up the point $p_0=[0,0,0,1]$.  We let $H$ denote the class that descends from the hyperplane of $\mathbb{P}^3$ and $E$ denote the exceptional divisor.  There are correspondingly two curve classes.  We use $\ell_0$ for the descendent of the intersection of hyperplanes in $\mathbb{P}^3$ and $\ell$ the nontrivial $\mathbb{P}^1$ inside $E$.  The intersection data in this blown-up space, which we refer to as $X$, is
\begin{equation}\begin{array}{c|cc}
& H & E \\ \hline
H & \ell_0 & 0 \\
E & 0 & -\ell
\end{array}\end{equation}
\begin{equation}\begin{array}{c|cc}
& H & E \\ \hline
\ell_0 & 1 & 0 \\
\ell & 0 & -1
\end{array}\end{equation}
In these computations, we used the fact that $(H-E)\cdot E = \ell$ and $(H-E)\cdot \ell=1$.  Nonvanishing triple intersections are
\begin{equation}H^3=1\qquad E^3=1\,.
\end{equation}

Within this blown-up space, we consider the proper transform of the nodal cubic curve
\begin{equation}{\cal{C}}:\qquad Z_3=0\,,\qquad Z_4Z_1Z_2+(Z_1+Z_2)^3=0 \,.
\end{equation}
A smooth cubic curve inside the $\mathbb{P}^2$ defined by $Z_3=0$ is a torus.  The nodal cubic here is a pinched torus which is topologically equivalent to a $\mathbb{P}^1$ glued to itself by identifying the north and south poles.  When we take the proper transform of ${\cal{C}}$, we separate the north and south poles and get an honest $\mathbb{P}^1$.  If we want to describe the proper transform in equations, we can do it as follows.  Each of the hyperplanes $Z_i=0$ for $i=1,2,3$ becomes irreducible, containing one component in the class $H-E$ and another in the class $E$.  Accordingly, we can write
\begin{equation}Z_i = \zeta W_i\,,\qquad i=1,2,3\end{equation}
for $W_i$ sections of the bundle ${\cal{O}}(H-E)$ and $\zeta$ a section of ${\cal{O}}(E)$ whose vanishing defines the exceptional divisor.  The equations for ${\cal{C}}$ now become
\begin{equation}\zeta W_3=0\,,\qquad \zeta^2\left(Z_4W_1W_2 + \zeta(W_1+W_2)^3\right)=0\,.\end{equation}
The proper transform is obtained by dropping the $\zeta$ prefactors.  Note that there are no solutions to $W_1=W_2=W_3=0$ since they are really just (global extensions of) projective coordinates on the exceptional $\mathbb{P}^2$.  This is what ensures that the curve
\begin{equation}{\cal{C}}':\qquad W_3=0\,,\qquad Z_4W_1W_2+\zeta(W_1+W_2)^3=0\end{equation}
is smooth.  As we said ${\cal{C}}'$ is a $\mathbb{P}^1$ so, when we blow up along ${\cal{C}}'$ to get our $B_3$, the exceptional divisor of that blow-up will be a Hirzebruch.  To see which one, note that ${\cal{C}}'$ is in the class
\begin{equation}
[{\cal{C}}']=(H-E)\cdot (3H-2E) \,.
\end{equation}
The normal bundle is a sum of line bundles and, because ${\cal{C}}'$ is a $\mathbb{P}^1$, these are determined by their degrees.  The normal bundle inside $(H-E)$ is computed by $(H-E)(3H-2E)^2=5$ while the normal bundle inside $3H-2E$ is computed by $(H-E)^2(3H-2E)=1$.  The normal bundle is therefore ${\cal{O}}(5)\oplus {\cal{O}}(1)$ so when we blow up along ${\cal{C}}'$ the exceptional divisor $D'$ is an $\mathbb{F}_4$.

The final step, as we said, is blowing up along ${\cal{C}}'$ to get the  threefold $B_3$ of \cite{Marsano:2009ym}.  We get an exceptional divisor $D'$ and let $E'$ denote the proper transform of $E$, which is a $dP_2$ surface.  We get a new curve class as well which is the $\mathbb{P}^1$ fiber of $D'$.  For our curve classes, we take $\ell_0$ to be the descendent of the curve of the same name in both $\mathbb{P}^3$ and $X$.  By $\ell$ we mean the proper transform of the curve of the same name ($\ell$) in $X$.  Finally, let us take the third curve class to be the fiber of our $\mathbb{F}_4$.  Because it is the fiber of a Hirzebruch surface, let us call this curve $f${\footnote{Note that $f$ is the curve that we called $\ell-G'$ in \cite{Marsano:2009ym}.}}

The intersection of divisors with curves is easy to compute
\begin{equation}\begin{array}{c|ccc}
& H & E' & D' \\ \hline
\ell_0 & 1 & 0 & 0 \\
\ell & 0 & -1 & 1 \\
f & 0 & 0 &  -1
\end{array}\end{equation}
The intersection of theHirzebruch $D'$ with $\ell$ is 1 because $\ell$ is the proper transform of a hyperplane in $E'$ (in the class $h-e_i$) while the Hirzebruch is a fibration of $f$ over the curve ${\cal{C}}'$.  The curve $f$ reduces to an exceptional curve in $E'$ which intersects $h-e_i$ exactly once.  The intersection of $D'$ with $f$ is computed by noting that $H-D'-E'$ meets $f$ exactly once.

We now turn to intersections of divisors.  Part of the intersection table is easy to compute
\begin{equation}\begin{array}{c|ccc}
& H & E' & H-D'-E'  \\ \hline
H & \ell_0 & 0 & \ell_0-3f \\
E' & 0 & \ast & \ast \\
H-D'-E' & \ell_0 - 3f & \ast & \ast
\end{array}\end{equation}

We can fill in the rest by noting some useful facts.  First, the intersection of $H-E$ with $E$ is just $\ell$ so the intersection of $H-E'$ with $E'$ is the total transform $\ell+f$.  The reason for this is that $H-E$ does not contain the curve ${\cal{C}}'$.  From this we conclude that
\begin{equation}(H-E')\cdot E' = \ell+f\quad \implies\quad E^{\prime\,2} = -(\ell+f) \,.
\end{equation}
Now, we know that $E'$ meets $D'$ in twice the curve class $f$ so we have that
\begin{equation}E'\cdot D' = 2f\quad\implies\quad (H-D'-E')\cdot E' = \ell-f  \,.
\end{equation}
Finally, we need $(H-D'-E')^2$.  For this, it will be enough to get $D^{\prime\,2}$.  We can discern this, however, from
\begin{equation}(H-D'-E')\cdot (3H-D'-2E')=0 \,.
\end{equation}
This leads to
\begin{equation}(H-D'-E')^2 = \ell+5f-2\ell_0\end{equation}
and we complete the intersection table as
\begin{equation}\begin{array}{c|ccc}
& H & E' & H-D'-E'  \\ \hline
H & \ell_0 & 0 & \ell_0-3f \\
E' & 0 & -\ell-f & \ell - f \\
H-D'-E' & \ell_0 - 3f & \ell - f & \ell+5f-2\ell_0 \\
D' & 3f & 2f & 3\ell_0 -7f - 2\ell
\end{array}\end{equation}

Now, for instantons we are particularly interested in divisors $S_2$ with trivial $H^0(S_2,N_{S_2/B_3})$.  These will be divisors that do not move in families inside $B_3$. The candidate divisors are
\begin{equation}\begin{split}E' &\cong dP_2 \\
D' &\cong \mathbb{F}_4 \\
H-D'-E' &\cong dP_1 \,.
\label{candidateS2s}\end{split}\end{equation}


\subsection{Divisors and Normal Bundles}

To study whether a divisor $S_2$ plays a role in the generation of nonperturbative couplings via M5 instantons, it is not enough to know that $h^0(S_2,N_{S_2/B_3}=0$.  We would also like to know $h^p(S_2,N_{S_2/B_3})$ for $p=1,2$.  In this subsection, we compute these cohomologies for the divisors in \eqref{candidateS2s}.

\subsubsection{$E'=dP_2$}

First we start with $E'=dP_2$.  Denote by $h$, $e_1$, and $e_2$ the standard generators of $H_2(dP_2, \mathbb{Z})$.  We have
\begin{equation}e_1\sim e_2 = f\,,\qquad h = \ell+f \,.
\end{equation}
Now, the normal bundle is given by
\begin{equation}E^{\prime\,2}= -\ell-f = -h \,.
\end{equation}
 We can compute the relevant cohomologies as follows
\begin{equation}\begin{split}H^0(dP_2,{\cal{O}}(-h)) &= 0 \\
H^2(dP_2,{\cal{O}}(-h)) = H^0(dP_2,{\cal{O}}(-2h+e_1+e_2)) 
&= 0 \,.
\end{split}\end{equation}
These vanishings mean that
\begin{equation}\begin{split}
h^1(dP_2,{\cal{O}}(-h)) &= -\chi(dP_2,{\cal{O}}(-h)) \\
&= 0 \,.
\end{split}\end{equation}
To do this computation, we use the fact that
\begin{equation}\begin{split}\chi(S,L) &= \int_S\,\text{ch}(V)\wedge\text{Td}(S) \\
&= \int_S \left[\frac{1}{12}\left(c_1(TS)^2+c_2(TS)\right) + \frac{1}{2}c_1(L)^2 + \frac{1}{2}c_1(L)c_1(TS)\right]
\end{split}\end{equation}
along with
\begin{equation}c_1(T_{dP_n}) = {\cal{O}}(3h-\sum_{i=1}^ne_i)\qquad \int_{dP_n}\left[c_2(T_{dP_n})+c_1(T_{dP_n})^2\right]=12 \,.
\end{equation}
To summarize, we find
\begin{equation}\boxed{h^{0,p}(E',N_{E'/B_3})=0}\end{equation}
Of course we needed this just to ensure that the 7-brane GUT theory did not contain chiral adjoints.

\subsubsection{$D'=\mathbb{F}_4$}
\label{app:subsubsecF4}

We now turn to $D'=\mathbb{F}_4$.  We usually use $b$ for the base of the Hirzebruch and $f$ for the fiber.  We know that the curve $f$ in $B_3$ is just the fiber of $\mathbb{F}_4$ by construction.  On the other hand, we expect that $b$ is given by the intersection of $D'$ with $(H-D'-E')$.  This is because $H-D'-E'$ is the proper transform of the hyperplane containing the $\mathbb{P}^1$ along which we blew up to get $D'$.  Intersecting $D'$ with this hyperplane should give the base of the Hirzebruch.  This means we have
\begin{equation}f\sim f \qquad b\sim (H-D'-E')\cdot D' = 3\ell_0-2\ell-7f\,.\end{equation}
We can check this identification by computing some intersections inside $D'$.  Using the relations
\begin{equation}D'\cdot E' = 2f\qquad D'\cdot (H-D'-E')=b\end{equation}
we find
\begin{equation}\begin{split}2f\cdot_{D'}f &= E'\cdot f \\
&= 0 \\
2f\cdot_{D'}b &= E'\cdot (3\ell_0-2\ell-7f) \\
&= 2 \\
b\cdot_{D'}b &= (H-D'-E')\cdot (3\ell_0-2\ell-7f) \\
&= -4 \,.
\end{split}\end{equation}
From this, we learn that
\begin{equation}f\cdot_{D'}f = 0\qquad f\cdot_{D'}b = 1\qquad b\cdot_{D'}b = -4 \,,
\end{equation}
which is precisely the intersection table of $\mathbb{F}_4$.

To move further, it is useful to know the Chern classes of $\mathbb{F}_4$.  Viewing $\mathbb{F}_n$ properly as a $\mathbb{P}^1$ fibration over $\mathbb{P}^1$ we have (recalling that $c(\mathbb{P}^1)=1+2f$)
\begin{equation}c(\mathbb{F}_n) = c(\mathbb{P}^1)(1+b)(1+b+nf)\end{equation}
so that
\begin{equation}c_1(\mathbb{F}_n) = 2b+(n+2)f\qquad c_2(\mathbb{F}_n)=4\end{equation}
Note that
\begin{equation}c_1(\mathbb{F}_n)^2=8\end{equation}

Now, the normal bundle of $D'$ is given by
\begin{equation}(D')^2 = 8f+2\ell-3\ell_0 = f-b\end{equation}
so what we want to compute is $h^{0,p}(\mathbb{F}_4,{\cal{O}}(f-b))$.  It is rather clear that
\begin{equation}H^0(\mathbb{F}_4,{\cal{O}}(f-b))=0\,.
\end{equation}
We also have
\begin{equation}H^2(\mathbb{F}_4,{\cal{O}}(f-b)) = H^0(\mathbb{F}_4,{\cal{O}}(b-f-(2b+6f))=0 \,.
\end{equation}
Now, we can evaluate
\begin{equation}\chi(\mathbb{F}_4,{\cal{O}}(f-b))= \frac{1}{2}(f-b)\cdot (b+7f) = 0 \,.
\end{equation}
This means that
\begin{equation}H^1(\mathbb{F}_4,{\cal{O}}(f-b)) = 0\,.
\end{equation}
To summarize, we find that
\begin{equation}\boxed{h^{0,p}(D',N_{D'/B_3}) = 0}\end{equation}

\subsubsection{$H-D'-E'=dP_1$}

Let us start by identifying the hyperplane $\tilde{h}$ and exceptional curve $\tilde{e}$ of $H-D'-E'$.  We have
\begin{equation}\tilde{h} = (H-D'-E')\cdot H = \ell_0-3f\qquad \tilde{e} = (H-D'-E')\cdot E' = \ell-f\end{equation}
To check this, we can compute
\begin{equation}\begin{split}\tilde{h}\cdot_{H-D'-E'}\tilde{h} &= H\cdot (\ell_0-3f) \\
&= 1 \\
\tilde{h}\cdot_{H-D'-E'}\tilde{e} &= H\cdot (\ell-f)\\
&= 0 \\
&= E'\cdot (\ell_0-3f) \\
&= 0 \\
\tilde{e}\cdot_{H-D'-E'}\tilde{e} &= E'\cdot (\ell-f) \\
&= -1 \,.
\end{split}\end{equation}
The normal bundle of $H-D'-E'$ is given by
\begin{equation}(H-D'-E')^2 = \ell+5f-2\ell_0 = -2\tilde{h}+\tilde{e} \,.
\end{equation}
Now it immediately follows that
\begin{equation}
\begin{split}
H^0(dP_1,{\cal{O}}(-2\tilde{h}+\tilde{e}) &= 0 \\
H^2(dP_1,{\cal{O}}(-2\tilde{h}+\tilde{e}) &= H^0(dP_1,{\cal{O}}(-h)) \\
&= 0 \,.
\end{split}\end{equation}
Now we compute
\begin{equation}\chi(dP_1,{\cal{O}}(-2\tilde{h}+\tilde{e}) = 1 + \frac{1}{2}(-2\tilde{h}+\tilde{e})\cdot (\tilde{h}) = 0
\end{equation}
to conclude that
\begin{equation}H^1(dP_1,{\cal{O}}(-2\tilde{h}+\tilde{e})) = 0 \,.
\end{equation}

So, to summarize, we have that
\begin{equation}\boxed{h^{0,p}(H-D'-E',N_{(H-D'-E')/B_3})=0}\end{equation}

\newpage

\renewcommand{\refname}{Bibliography}
\addcontentsline{toc}{section}{Bibliography}

\providecommand{\href}[2]{#2}\begingroup\raggedright\endgroup

\end{document}